\documentclass[twoside,twocolumn,9pt]{article}
\usepackage{extsizes}
\usepackage[super,sort&compress,comma]{natbib} 
\usepackage[version=3]{mhchem}
\usepackage[left=1.5cm, right=1.5cm, top=1.785cm, bottom=2.0cm]{geometry}
\usepackage{balance}
\usepackage{mathptmx}
\usepackage{sectsty}
\usepackage{graphicx} 
\usepackage{lastpage}
\usepackage[format=plain,justification=justified,singlelinecheck=false,font={stretch=1.125,small,sf},labelfont=bf,labelsep=space]{caption}
\usepackage{float}
\usepackage{fancyhdr}
\usepackage{fnpos}
\usepackage[english]{babel}
\addto{\captionsenglish}{%
}
\usepackage{array}
\usepackage{droidsans}
\usepackage{charter}
\usepackage[T1]{fontenc}
\usepackage[usenames,dvipsnames]{xcolor}
\usepackage{setspace}
\usepackage[compact]{titlesec}
\usepackage{hyperref}

\usepackage[ruled]{algorithm2e}

\usepackage{etoolbox}
\AtBeginEnvironment{algorithm}{\SetArgSty{textrm}}

\SetStartEndCondition{ }{}{}
\SetKwProg{Fn}{def}{\string:}{}
\SetKwFunction{Range}{range}
\SetKwFor{For}{for}{\string:}{}
\SetKwIF{If}{ElseIf}{Else}{if}{:}{elif}{else:}{}
\SetKwFor{While}{while}{:}{}
\AlgoDontDisplayBlockMarkers\SetAlgoNoEnd\SetAlgoNoLine

\newcommand{\DefineAuthor}[2]{
  \expandafter\newcommand\csname #1note\endcsname[1]{
    \textbf{\textcolor{#2}{#1: ##1}}}
  \expandafter\newcommand\csname #1\endcsname[1]{
    \textcolor{#2}{##1}}
  \expandafter\newcommand\csname #1cancel\endcsname[1]{
    \textcolor{#2}{\sout{##1}}}
  \expandafter\newcommand\csname #1change\endcsname[2]{
    \textcolor{#2}{\sout{##1} ##2}}
  \newenvironment{#1text}{\color{#2}}{\color{black}}
}
\usepackage{xcolor}
\definecolor{GPmagenta}{rgb}{1.0, 0.0, 1.0}
\definecolor{GPgreen}{rgb}{0.0, 0.5019607843137255, 0.0}

\usepackage{comment}

\usepackage{MnSymbol}
\usepackage{physics}
\usepackage{wasysym}

\definecolor{cream}{RGB}{222,217,201}

\begin{document}

\pagestyle{fancy}
\thispagestyle{plain}
\fancypagestyle{plain}{
\renewcommand{\headrulewidth}{0pt}
}

\makeFNbottom
\makeatletter
\renewcommand\LARGE{\@setfontsize\LARGE{15pt}{17}}
\renewcommand\Large{\@setfontsize\Large{12pt}{14}}
\renewcommand\large{\@setfontsize\large{10pt}{12}}
\renewcommand\footnotesize{\@setfontsize\footnotesize{7pt}{10}}
\makeatother

\renewcommand{\thefootnote}{\fnsymbol{footnote}}
\renewcommand\footnoterule{\vspace*{1pt}%
\color{cream}\hrule width 3.5in height 0.4pt \color{black}\vspace*{5pt}} 
\setcounter{secnumdepth}{5}

\makeatletter 
\renewcommand\@biblabel[1]{#1}            
\renewcommand\@makefntext[1]%
{\noindent\makebox[0pt][r]{\@thefnmark\,}#1}
\makeatother 
\renewcommand{\figurename}{\small{Fig.}~}
\sectionfont{\sffamily\Large}
\subsectionfont{\normalsize}
\subsubsectionfont{\bf}
\setstretch{1.125} 
\setlength{\skip\footins}{0.8cm}
\setlength{\footnotesep}{0.25cm}
\setlength{\jot}{10pt}
\titlespacing*{\section}{0pt}{4pt}{4pt}
\titlespacing*{\subsection}{0pt}{15pt}{1pt}

\fancyfoot{}
\fancyfoot[LO,RE]{\vspace{-7.1pt}\includegraphics[height=9pt]{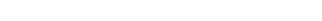}}
\fancyfoot[CO]{\vspace{-7.1pt}\hspace{13.2cm}\includegraphics{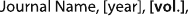}}
\fancyfoot[CE]{\vspace{-7.2pt}\hspace{-14.2cm}\includegraphics{head_foot/RF}}
\fancyfoot[RO]{\footnotesize{\sffamily{1--\pageref{LastPage} ~\textbar  \hspace{2pt}\thepage}}}
\fancyfoot[LE]{\footnotesize{\sffamily{\thepage~\textbar\hspace{3.45cm} 1--\pageref{LastPage}}}}
\fancyhead{}
\renewcommand{\headrulewidth}{0pt} 
\renewcommand{\footrulewidth}{0pt}
\setlength{\arrayrulewidth}{1pt}
\setlength{\columnsep}{6.5mm}
\setlength\bibsep{1pt}

\makeatletter 
\newlength{\figrulesep} 
\setlength{\figrulesep}{0.5\textfloatsep} 

\newcommand{\topfigrule}{\vspace*{-1pt}%
\noindent{\color{cream}\rule[-\figrulesep]{\columnwidth}{1.5pt}} }

\newcommand{\botfigrule}{\vspace*{-2pt}%
\noindent{\color{cream}\rule[\figrulesep]{\columnwidth}{1.5pt}} }

\newcommand{\dblfigrule}{\vspace*{-1pt}%
\noindent{\color{cream}\rule[-\figrulesep]{\textwidth}{1.5pt}} }

\makeatother

\twocolumn[
  \begin{@twocolumnfalse}
{\includegraphics[height=30pt]{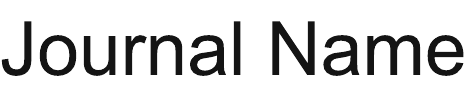}\hfill\raisebox{0pt}[0pt][0pt]{\includegraphics[height=55pt]{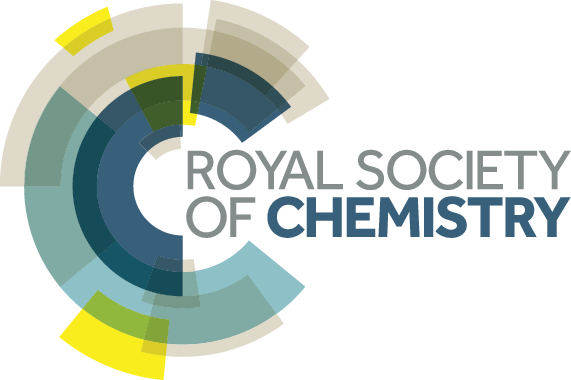}}\\[1ex]
\includegraphics[width=18.5cm]{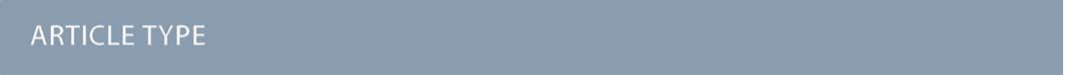}}\par
\vspace{1em}
\sffamily
\begin{tabular}{m{4.5cm} p{13.5cm} }

\includegraphics{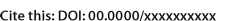} & \noindent\LARGE{\textbf{Quantum State Preparation Of Multiconfigurational States For Quantum Chemistry
}} \\
\vspace{0.3cm} & \vspace{0.3cm} \\

 & \noindent\large{
 Gabriel Greene-Diniz, 
 Georgia Prokopiou, 
 David Zsolt Manrique, 
 and David Mu{\~n}oz Ramo
 } \\

\includegraphics{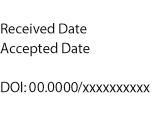} & \noindent\normalsize{The ability to prepare states for quantum chemistry is a promising feature of quantum computers, and efficient techniques for chemical state preparation is an active area of research. In this paper, we implement and investigate two methods of quantum circuit preparation for multiconfigurational states for quantum chemical applications. It has previously been shown that controlled Givens rotations are universal for quantum chemistry. To prepare a selected linear combination of Slater determinants (represented as occupation number configurations) using Givens rotations, the gates that rotate between the reference and excited determinants need to be controlled on qubits outside the excitation (external controls), in general. We implement a method to automatically find the external controls required for utilizing Givens rotations to prepare multiconfigurational states on a quantum circuit. We compare this approach to an alternative technique that exploits the sparsity of the chemical state vector and find that the latter can outperform the method of externally controlled Givens rotations; highly reduced circuits can be obtained by taking advantage of the sparse nature (where the number of basis states is significantly less than 2$^{n_q}$ for $n_q$ qubits) of chemical wavefunctions. We demonstrate the benefits of these techniques in a range of applications, including the ground states of a strongly correlated molecule, matrix elements of the Q-SCEOM algorithm for excited states, as well as correlated initial states for a quantum subspace method based on quantum computed moments and quantum phase estimation.} \\

\end{tabular}

 \end{@twocolumnfalse} \vspace{0.6cm}

  ]

\renewcommand*\rmdefault{bch}\normalfont\upshape
\rmfamily
\section*{}
\vspace{-1cm}


\footnotetext{\textit{~Quantinuum Ltd., 13-15 Hills Road, CB2 1NL Cambridge, United Kingdom; \\ E-mail: gabriel.greene-diniz@quantinuum.com}}




\section{Introduction}

Quantum computation has the potential to make a large impact on quantum chemistry and condensed matter physics. This has motivated a recent rapid development of techniques for representing quantum chemical states on gate-based quantum computers \cite{anselmetti21, yordanov20, yordanov21, tilly21, ryabinkin18, evangelista19, sugisaki19, grimsley20, anand22, chee23, arrazola22, fomichev24}. From the point of view of general quantum state preparation, many research works have focused on maximizing the efficiency of loading classical data into a quantum processor and preparing a quantum state to represent those data \cite{grover02, mottonen05, shende06, araujo21, zhang22, deveras22}. It is interesting to consider applications of the latter approach to quantum chemistry, where the classical data represent previously obtained chemical information, and the quantum computer can be used to further evolve the quantum chemical state (for example, to increase the accuracy of the representation of the ground or excited states of a molecule).

The existence of gate sets that are universal for quantum chemistry (in the sense of spanning the space of all states that preserve fermionic symmetries) has recently been proved \cite{arrazola22}. These gates take the form of Givens rotations (GRs) to represent particle-number-conserving excitations that correlate the occupations of different fermionic modes. GRs have also been used to construct gate fabrics that can navigate fermionic Fock spaces \cite{anselmetti21}. Regarding the preparation of multiconfigurational states (where each configuration represents electrons distributed throughout a molecular orbital basis, in a second quantized framework), if the configurations are specified \textit{a priori} then a sequence of GRs can be applied to the circuit to prepare the desired state vector \cite{arrazola22}. However, GRs will mix any basis states with qubit subspaces that match those involved in the rotation. To ensure that only the desired basis states are included in the state vector, GRs can be controlled by qubits outside the rotation space \cite{arrazola22} (which we refer to as ``external'' controls to distinguish these from internal CNOTs within the GR gate decomposition). However, externally controlling each GR on all qubits not directly involved in the rotation leads to very large circuits when decomposed into typical hardware gate sets. The question then remains as to how to apply external controls to guarantee the desired state vector is produced, while avoiding, when possible, externally controlling all GRs on all qubits outside the excitation. A specific example of this was presented in \cite{arrazola22} for a 6-qubit state consisting of 4 occupation number (ON) configurations. In this paper, we provide a general algorithm for finding external GR controls automatically for any particle-number-preserving fermionic state. 

We also consider the previously proposed method of Gleinig and Hoefler to prepare sparse quantum states \cite{gleinig21}. Initially designed to overcome the problem of exponentially scaling resources for arbitrary quantum state preparation \cite{mottonen05, plesch11, niemann16}, the sparse state preparation (SSP) method takes advantage of the fact that many interesting quantum states span a relatively small section of the full Hilbert space, where the number of non-zero basis coefficients is far less than $O(2^n)$. This is the case for typical states of interest in quantum chemistry; the full configuration interaction (FCI) \cite{pink_book_ch1} wavefunction obeys fundamental fermionic symmetries (e.g. spin and particle number conservation) which rule out many basis states of the entire Hilbert space. Additionally, the success of methods such as semistochastic heat-bath configuration interaction (SHCI) \cite{li18} and perturbatively selected configuration interaction (CIPSI) \cite{dash18} exemplify the fact that a relatively small number of configurations can often provide a good approximation to the full wavefunction.

The importance of the sparsity of chemical wavefunctions is emphasized when considering approximations to the ground state, particularly in situations requiring ``warm start'' states, i.e. those states with large overlap with the true ground state (relative to single configuration Hartree-Fock (HF) states) yet remain sparse enough to be efficiently prepared. Considering the example of quantum phase estimation (QPE) in which the success probability is proportional to the overlap between the initial state and the true ground state, and the fact that in many interesting cases (e.g. strongly correlated systems) the HF state has significantly low overlap, the ability to efficiently and conveniently prepare multiconfigurational states can be highly beneficial. A recently published work \cite{feniou24} noted the method of Gleinig and Hoefler as a non-variational approach to quantum chemical state preparation, and provided a comparison to other sparse state preparation approaches in terms of asymptotic scaling. Here, an implementation of Gleinig and Hoefler's approach suitable for quantum chemistry, along with its integration with various ground and excited state methods, is reported and applied to a strongly correlated chemical system.

In this paper we compare the techniques of externally controlled GRs \cite{arrazola22} and Gleinig and Hoefler's scheme \cite{gleinig21} for preparing quantum circuits corresponding to selected linear combinations of computational basis states, where the latter represent occupation number configurations (Slater determinants). These techniques are implemented in the InQuanto \cite{inquanto_web, inquanto_medium, inquanto_docs} software package, and we demonstrate their utility for a wide range of quantum algorithms useful for quantum chemistry.

\section{Preparation of Multiconfigurational States in Quantum Circuits}\label{methods}
\subsection{Givens rotations with external controls}\label{sec:gr_method}

Consider an arbitrary single electronic configuration specified by an ON configuration \cite{pink_book_ch1} in a finite basis of spin orbitals $|..., f_i, ..., f_j, ...\rangle$ with fermionic occupation numbers $f_i, f_j \in \{0, 1\}$, where $f_i \neq f_j$, and $i, j$ label orbitals involved in the excitation. An excitation can be viewed as a mixing between ON configurations with different occupations. Adopting the Jordan-Wigner (JW) transformation between fermionic and qubit algebras \cite{jordan28, szalay21} $|..., f_i, ..., f_j, ...\rangle \mapsto |..., q_i, ..., q_j, ...\rangle$ where $q_i, q_j \in \{0, 1\}$ are individual qubit states, a fermionic excitation can be encoded as a unitary operation applied to a register of qubits representing the ON configuration, rotating it to a convex combination
\begin{equation} \label{eqn:ex_as_ugate}
    \begin{split}
        &\mathcal{C}|..., f_i, ..., f_j, ...\rangle \mapsto \mathcal{U}|..., q_i, ..., q_j, ...\rangle \\
        &= c|..., q_i, ..., q_j, ...\rangle + \sqrt{1 - c^2}|..., q_j, ..., q_i, ...\rangle .
    \end{split} 
\end{equation}
\noindent Here, $\mathcal{C}$ is a fermionic excitation operator, and $\mathcal{U}$ is a unitary operator (encoded as quantum gates) applied to the qubit register. Note that $\mathcal{C}$ ($\mathcal{U}$) should preserve the total number of electrons (the qubit register Hamming weight). 

Eq. \ref{eqn:ex_as_ugate} corresponds to a 1-body excitation. Using a 2-qubit state as a minimal example, the excitation can be expressed in the computational basis as a GR \cite{anselmetti21, arrazola22, chee23}, whose action on the $2^{n_q=2}$ dimensional Hilbert space is to rotate in a 2 dimensional subspace of the state vector

\begin{equation} \label{eqn:2qGRsv}
    \begin{pmatrix} 
        1 & 0 & 0 & 0 \\ 
        0 & c_1 & c_2 & 0 \\
        0 & c_3 & c_4 & 0 \\
        0 & 0 & 0 & 1
    \end{pmatrix}
    \left(
    \begin{matrix} 
         & |00\rangle, & \\ 
         & |01\rangle, & \\
         & |10\rangle, & \\
         & |11\rangle, &
    \end{matrix}
    \right) = \left(
    \begin{matrix} 
         & |00\rangle, & \\ 
         & c_1|01\rangle+c_2|10\rangle, & \\
         & c_3|01\rangle+c_4|10\rangle, & \\
         & |11\rangle, &
    \end{matrix}
    \right).
\end{equation}
\noindent The coefficients of the orthonormal basis are omitted for the sake of clarity (the state vector is normalized to unity). The real scalars are restricted by $|c_1^2| + |c_2^2| = |c_3^2| + |c_4^2| = 1$ with $c_2$ and $c_3$ having opposite signs, to ensure unitarity. This GR can be written as
\begin{equation}\label{eqn:2qGR}
    \mathcal{G}^2(\theta) = 
    \begin{pmatrix} 
        1 & 0 & 0 & 0 \\ 
        0 & \cos{\theta} & \sin{\theta} & 0 \\
        0 & -\sin{\theta} & \cos{\theta} & 0 \\
        0 & 0 & 0 & 1
    \end{pmatrix}
\end{equation}
\noindent where $\theta$ parameterizes the quantum gates implementing the Givens rotation \cite{arrazola22, anselmetti21}. Eqs. \ref{eqn:2qGRsv} and \ref{eqn:2qGR} show, in a small example, that individual GRs preserve the particle number within the subspace to which they are applied. Generalizations up to $n_{\text{elec}}$-body excitations (with $n_{\text{elec}}$ the number of electrons) for $n_q \geq 2n_{\text{elec}}$ qubits are obtained by extending the mixing elements to other entries of the $2^{n_q} \times 2^{n_q}$ matrix (explicit matrices for 2-body rotations on 4 qubits are given in \cite{anselmetti21}). 

The Hamming distance between two binary strings $x$ and $y$, defined as

\begin{equation}\label{ham_sum}
    \mathfrak{h}(x, y) = \sum_i^{n_q} x(i) \oplus y(i),
\end{equation}

\noindent plays a central role in the use of GRs to prepare multiconfigurational states. $\mathfrak{h}$ is related to the excitation order $\lambda = \mathfrak{h}/2$ for $\lambda$-body excitations between two ON configurations. The qubits on which the excitation unitary acts can be specified using $\mathfrak{h}$ and its decomposition into an array of qubit-wise XOR values $(x(0) \oplus y(0), x(1) \oplus y(1), ..., x(n_q) \oplus y(n_q))$. The indexes of non-zero elements of this array correspond to the GR qubit indexes, while the value of $\mathfrak{h}$ specifies the excitation order. 

Selected linear combinations of ON configurations with fixed coefficients can be prepared using sequences of GRs applied to the circuit. Each GR mixes a pair of basis states (\textit{i.e.} excitation) and its rotation angle $\theta$ can be obtained from state vector coefficients via recursive normalization (see Arrazola \textit{et al.}\cite{arrazola22}). Alternatively, optimized coefficients can be found by varying the gate angles within each rotation in a variational algorithm. As mentioned in the introduction (and elaborated in previous work \cite{arrazola22}), the gates that encode each GR may need to be externally controlled to ensure that the sequence of GRs does not yield unwanted basis states. Thus, all the steps necessary to prepare a linear combination of ON configurations on a quantum circuit are at hand, apart from a general method to specify the external controls. 

In the following subsection, we outline a procedure to automatically accomplish this task given a selected set of ON configurations, which specifies the excitation unitaries and the external controls needed to accomplish each GR. This can be used to build the circuit representing the multiconfigurational state.

\subsubsection{Algorithm to specify Givens rotations and external controls}\label{}

In Algorithm \ref{alg:gr}, a procedure is presented to find a sequence of GRs and their external controls required to prepare a selected linear combination of ON configurations. The input is an ordered set of $D$ bit strings representing the ON configurations. Note that the reference for excitations is the first member of the input set, and the desired state vector is built up as sequential excitations relative to this reference. The ordered $D - 1$ Hamming distances $\mathfrak{h}(x_1, x_{d>1})$ and their qubit components are then stored. Gate angles of each GR, dependent on the ON coefficients, can be found by using recursive normalization of ON coefficients (a specific example is also discussed in sec. \ref{sec:res_4qvqe}).

Let $\mathcal{G}_e$ represent a rotation that linearly mixes the reference with a basis state indexed by $e=2, \dots, D$. The excitation indexes are obtained from a decomposition of the Hamming distance (non-zero contributions to the sum in Eq. \ref{ham_sum}), which specifies each GR unitary in the ($D - 1$)-length sequence, \textit{i.e.} $\mathcal{G}_e^{\mathfrak{h}(x_1, x_e)} = \mathcal{G}(\textbf{\textit{i}}^{\mathfrak{h}(x_1, x_e)}_{e})$, where $\textbf{\textit{i}}^{\mathfrak{h}}_{e}$ is a tuple of indexes of length $2\lambda$ corresponding to nonzero contributions to $\mathfrak{h}(x_1, x_{e})$. Note that for the basis state index $e=2$, external controls are not needed by definition.

For a given ordered set of input configurations, the search for external controls required for $\mathcal{G}^{\mathfrak{h}(x_1, x_e)}_{e}$ depends on whether a ``previous'' ($p < e$) basis state is ``rotatable'' by $\mathcal{G}^{\mathfrak{h}(x_1, x_e)}_{e}$, and if so controls are applied to ensure that $\mathcal{G}^{\mathfrak{h}(x_1, x_e)}_{e}$ will not act on any basis state with index $p=2, ..., e-1$. 

For basis states $x_e$ separated from the reference $x_1$ by $\mathfrak{h}(x_1, x_e) = 2$ or $\mathfrak{h}(x_1, x_e) = 4$, Givens rotations $\mathcal{G}^2_{e}$ or $\mathcal{G}^4_{e}$, respectively, can be used to mix $x_e$ with $x_1$, and the external controls found by Algorithm \ref{alg:gr} ensure that the circuit maintains the desired state vector. For $\mathcal{G}^2_{e}$ and $\mathcal{G}^4_{e}$, we use the gate decompositions presented in previous works for 1-body and 2-body GRs \cite{anselmetti21, arrazola22} (compilations of these circuits in the H-series gate set \cite{h11} are given in the Appendix Fig. \ref{fig:gr2_gr4}). However, when $\mathfrak{h}(x_1, x_e) > 4$, a unitary is required that encodes the ($\lambda>2)$-body excitation. To this end, we adapt a gadget described by Arrazola \textit{et al.} \cite{arrazola22} consisting of a) a sequence of multi-controlled SWAPs, b) a central $\mathcal{G}^1_{e}$, and c) the reverse of a). In the appendix Sec. \ref{app:sec1}, Algorithm \ref{alg:lambda_gt_2} describes how the indexes of these unitaries and their external controls are found. Compared to Arrazola \textit{et al.} \cite{arrazola22}, our scheme achieves a lower overhead of external controls for SWAPs and the central $\mathcal{G}^2_{e}$ to combine ON configurations separated by $\lambda>2$. An example of this for a 3-body excitation of the $\ket{111000}$ configuration is shown in Fig. \ref{fig:triple}.

Once the $\mathcal{G}^{\mathfrak{h}(x_1, x_e)}_{e}$ for all $x_e$ are found, along with their external controls (if required), the desired state $|\psi\rangle$ can be prepared as

\begin{equation}\label{eqn:gr_psi}
  |\psi\rangle = \bigg(\prod_e \mathcal{G}_e^{\mathfrak{h}(x_1, x_e)} \bigg) |x_1\rangle \ .
\end{equation}

\begin{algorithm}[hbt!]
\caption{State Preparation with GRs}\label{alg:gr}
\textbf{Input:} Ordered set of bit strings (ON configurations) $(x_1, ...,  x_{D})$ representing $|\psi\rangle = \sum_{d=1}^Dc_d|x_d\rangle$, each with binary values $x_d(i)$ on the $i^{\text{th}}$ qubit. \\

\textbf{Output:} Sequence of qubit indexes of $D-1$ GRs and their external controls. \\

\nl Collect $D-1$ GR indexes $\{\textbf{\textit{i}}^{\mathfrak{h}(x_1, x_e)}_e\}$ for $e = 2, ..., D$.\;

\nl \For {$e=3$ \KwTo $D$}{
\nl    \If{$\mathfrak{h}(x_1, x_e) = 2 $}{
\nl        \For{$p=2$ \KwTo $e-1$}{
\nl            \If{$x_p(i \in \textbf{\textit{i}}^2_e) \neq x_p(j \in \textbf{\textit{i}}^2_e)$}{
\nl                Set external control index of $\mathcal{G}(\textbf{\textit{i}}^2_e)$ to lowest $i_p$ such that $x_p(i_p) \neq x_e(i_p)$ and $i_p \notin \textbf{\textit{i}}^2_e$, with control state $x_1(i_p)$\;
            }
        }
    }
\nl    \If{$\mathfrak{h}(x_1, x_e) = 4 $}{
\nl        \For{$p=2$ \KwTo $e-1$}{
\nl             $\mathfrak{h}(x, x_p; \textbf{\textit{i}}^4_e) = \sum_{i \in \textbf{\textit{i}}^4_e} x(i) \oplus x_p(i)$\;
\nl             \If{\normalfont $\mathfrak{h}(x_1, x_p; \textbf{\textit{i}}^4_e) = 4$ or $\mathfrak{h}(x_e, x_p; \textbf{\textit{i}}^4_e) = 4$}{
\nl                 Set external control index of $\mathcal{G}(\textbf{\textit{i}}^4_e)$ to lowest $i_p$ such that $x_p(i_p) \neq x_e(i_p)$ and $i_p \notin \textbf{\textit{i}}^4_e$, with control state $x_1(i_p)$\;
            }
        }
    }
\nl    \If{$\mathfrak{h}(x_1, x_e) > 4 $}{
\nl        Use Algorithm \ref{alg:lambda_gt_2} with current value of $e$\;
    }
}
\nl Return ordered sequence of GR indexes and their external controls\; 
\end{algorithm}

\subsection{Sparse State Preparation for Quantum Chemistry}\label{sec:ssp_method}

In this section, we briefly describe Gleinig and Hoefler's sparse state preparation (SSP) method . For more details, we refer the reader to the original article \cite{gleinig21}. Here, we provide a brief description for completeness and highlight those aspects most relevant to multiconfigurational state preparation for quantum chemistry. This method essentially solves the problem of finding some $n$-qubit circuit $\mathcal{U^{\dagger}_{\text{SSP}}}$ such that $\mathcal{U^{\dagger}_{\text{SSP}}}|\psi \rangle = |0^{\otimes n} \rangle$, where $|\psi \rangle$ is the desired $n$-qubit state to be prepared. Knowing $\mathcal{U^{\dagger}_{\text{SSP}}}$, one can reverse this to obtain the state preparation unitary $\mathcal{U_{\text{SSP}}}$ such that the desired state is $|\psi \rangle = (\mathcal{U^{\dagger}_{\text{SSP}}})^{\dagger}|0^{\otimes n} \rangle = \mathcal{U_{\text{SSP}}}|0^{\otimes n} \rangle $. The input to the SSP method is an unordered set $ \mathcal{S} = \{x_1, ..., x_{D}\}$ of bit strings representing ON configurations, with normalized coefficients $c_d$ such that the state to be prepared is $|\psi(\mathcal{S}) \rangle = \sum_{d=1}^Dc_d|x_d\rangle$ and $\sum_d^D|c_d|^2 = 1$.

In the $i^{\text{th}}$ iteration of the algorithm \cite{gleinig21}, an arbitrary angle (possibly controlled) 1-qubit rotation gate along with an arrangement of NOT and CNOT gates (collectively labeled here as $\mathcal{U}_i$) are found, which if applied to the state at the $i^{\text{th}}$ iteration results in a new state hosting $|\mathcal{S}_i| =  D_i$ < $D_{i-1}$ non-zero basis coefficients (where $D_0 \equiv D = |\mathcal{S}|)$. This is accomplished by ``merging'' two bit strings $x_{1,i}, x_{2,i}$ contained in $\mathcal{S}_i$, where merging amounts to applying $\mathcal{U}_i$ such that $\mathcal{U}_i|\psi(\{..., x_{1,i}, x_{2,i}, ..., \})\rangle = |\psi'(\{..., x_{2,i}, ..., \})\rangle$ (up to global phase). The arrangement of gates in each $\mathcal{U}_i$ are determined by the distribution of binary values throughout the $D$ bit strings \cite{gleinig21}. We note that for real-valued $c_d$ the $R_y$$(\theta)$ gate parameterized by angle $\theta$ is sufficient to represent the 1-qubit rotation in each $\mathcal{U}_i$. 

The gate angles of the SSP method, which determine the values of $c_d$, are obtained on-the-fly and enter in the 1-qubit rotations,  each of which mixes 2 basis states $x_{1,i}, x_{2,i}$ at a given iteration $i$ of the SSP algorithm; the $i^{\text{th}}$ rotation angle is obtained by normalizing the $x_{2,i}$ coefficient $c_{2,i}$ to $\sqrt{c_{1,i}^2+c_{2,i}^2}$ and taking the arcsin. For the next $i+1$ iteration, the coefficient $c_{2,i}$ is replaced by the previous normalizer $\sqrt{c_{1,i}^2+c_{2,i}^2}$, and the angle of the $(i+1)^{\text{th}}$ single qubit rotation is obtained for $x_{1,i+1}, x_{2,i+1}$ as in the previous iteration. For an example of this, see Gleinig and Hoefler's paper \cite{gleinig21}.

An important consequence of this algorithm is that the sequence of $\mathcal{U}_i$ applied to the initial circuit $|0^{\otimes n} \rangle$ is independent of the order of bit strings provided by the user at input. This is a major difference with the GR method described in Sec. \ref{sec:gr_method}, and highlights the lack of a ``reference'' ON configuration for excitations in the SSP method (whereas in the GR method the excitation reference is fixed as the first member of the ordered set of input bit strings, and all GRs subsequently applied can be considered as rotations relative to the reference). While this does not affect the behavior of SSP-prepared circuits for fixed (non-variational) $c_d$, variational applications of this method (in which gate angles are variationally optimized, hence leading to variationally optimized $c_d$) can be affected, as discussed in Sec. \ref{sec:res_4qvqe}.

\subsection{Methods which benefit from Multiconfigurational State Preparation}

In the following subsections we mention a non-exhaustive list of methods which we use to demonstrate the utility of circuit preparation to represent selected linear combinations of ON configurations. We also mention previous applications, namely the preparation of $N\pm1$-particle states for Green's functions \cite{GreeneDiniz24}, and quantum circuit representations of CASSCF (complete active space self-consistent field) wavefunctions to study orbital entanglement \cite{orb_paper_scirep}. 

\subsubsection{Variational Quantum Eigensolver}\label{sec:vqe}
The variational quantum eigensolver (VQE) \cite{tilly21} is a hybrid quantum-classical algorithm. As initially proposed, VQE utilizes a quantum computer to perform measurements of the expectation value of a Hamiltonian taken with respect to a parameterized ansatz for the wavefunction, which is optimized by a classical processor based on a minimization of the energy according to the variational principle. 

On present-day noisy intermediate scale quantum (NISQ) hardware, the originally proposed form of VQE is significantly limited by various sources of noise (e.g. qubit decoherence, gate error rates, state preparation and measurement errors). Hence, in this work, VQE is utilized as a classical optimizer of gate angles $\vec{\theta}$ in multiconfigurational state circuits, which directly translates into optimization of basis state (ON configuration) coefficients of the circuit state vector. This demonstrates how the methods for preparing multiconfigurational state circuits (Sec. \ref{sec:gr_method} and Sec. \ref{sec:ssp_method}) can be used to prepare variational ansatzes for VQE applied to quantum chemistry. In this context, VQE is run in an ideal setting in which quantum measurements are classically simulated so that $|\psi(\vec{\theta})\rangle$ consists of ideally optimized basis coefficients. $|\psi(\vec{\theta})\rangle$ can then be used as the initial (warm start) state in subsequent methods (such as quantum subspace methods, or quantum phase estimation). One can also measure expectation values of the Hamiltonian in a noisy setting using offline optimized $|\psi(\vec{\theta})\rangle$ to quantum compute estimates of the ground state energy corresponding to ideal VQE parameters (see Sec. \ref{sec:res_4qvqe}).

\subsubsection{Quantum Computed Moments}\label{sec:qsm}

A wide range of quantum subspace methods currently exist \cite{motta24}, which generally attempt to solve for accurate approximations to low-lying eigenvalues of a given Hamiltonian $\hat{H}$ by projecting onto a finite-dimensional subspace spanned by a set of basis states. Krylov basis states correspond to a subspace spanned by applying powers of the Hamiltonian to a trial state (reference), and various forms of Krylov subspace methods have recently been investigated (see a recent review \cite{motta24} and references therein). The quantum computed moments (QCM) approach, recently introduced by Vallury \textit{et al.} \cite{vallury20} and subsequently applied in various contexts \cite{jones22, GreeneDiniz24}, also utilizes powers (or moments) of the Hamiltonian $\hat{H}$. While essentially constituting a subspace method, the QCM approach does not explicitly construct the Krylov basis states on a quantum circuit. Rather, QCMs provide a correction to the energy that encodes electronic correlation beyond that included in $\langle \hat{H} \rangle$ (where $\langle \rangle$ denotes the expectation value with respect to an input state that approximates the exact ground state). Based on the cumulant expansion of the Lanczos diagonalization method, the elements of the tridiagonalized representation of $\hat{H}$ can be expressed in terms of polynomials of $\langle \hat{H}^m \rangle$, where the maximum value of $m$ is related to the dimension $l$ of the subspace expansion as $2l - 1$. 

By deriving a truncation of the Lanczos expansion at $m=4$ \cite{hollenberg94}, a non-perturbative approximation to the ground state energy referred to as QCM4 can be expressed using cumulants $\mathfrak{c}_m$ (also referred to as connected moments)

\begin{equation}\label{eqn:cm}
\mathfrak{c}_m = \langle \hat{H}^m \rangle - \sum_{p = 0}^{m - 2} \binom{m - 1}{p} \mathfrak{c}_{p + 1} \langle \hat{H}^{m - p - 1} \rangle \ ,
\end{equation}

\begin{equation}\label{eqn:qcm4}
E_{\text{QCM4}} = \mathfrak{c}_1 - \mathfrak{c}_2\frac{\mathfrak{c}_2^2}{\mathfrak{c}_2^3 - \mathfrak{c}_2\mathfrak{c}_4} \left[ \sqrt{3\mathfrak{c}_3^2 - 2\mathfrak{c}_2\mathfrak{c}_4} - \mathfrak{c}_3 \right] \ .
\end{equation}

In this work, we also utilize the connected moments expansion \cite{cioslowski87, knowles87, kowalski20, claudino21} in which the ``t-expansion'' of Horn and Weinstein \cite{horn84} can be expressed in terms of $\mathfrak{c}_n$. This leads to alternative expressions of the energy to various orders of moments. Here we consider the CMX2 formula, in which ``2'' refers to the dimension of the associated subspace $l$

\begin{equation}\label{eqn:cmx2}
    E_{\text{CMX2}} = \mathfrak{c}_1 - \frac{\mathfrak{c}_2^2}{\mathfrak{c}_3}
\end{equation}

\noindent hence this can be considered as an energy correction up to $m=2l-1=3$. We compare CMX2 to QCM4 as a lower-cost (generally less measurements due to lower-order moments), yet more approximate form of the energy. 

To obtain expectation values $\langle \hat{H}^m \rangle$, the fermionic Hamiltonian is JW encoded as $\hat{H}^{m=1} = \sum_r h_{r,m=1} \hat{P}_{r,m=1}$ (where Pauli strings $\hat{P}_{r,m}$ are tensor products of Pauli gates spanning the qubit register and $h_{r,m}$ are real coefficients), then $\hat{H}^{m>1}$ are obtained by recursive multiplication of $\hat{H}$. $\langle \hat{H}^m \rangle$ can then be computed from quantum hardware measurements in the computational basis via Pauli string averaging. For ideal simulations, expectation values are computed classically (considered as the limit of infinite sampling in the absence of device noise). 

We note that the use of multiconfigurational (or multireference) initial states for connected moment expansions of the energy has a long history in classical computational chemistry \cite{cioslowski87}. In Sec. \ref{sec:res_8qqcm} we show a quantum computational version of this approach, where the circuit corresponding to the initial state is tailored to linear combinations of specific ON configurations (Slater determinants). 

\subsubsection{Quantum Phase Estimation and Hamiltonian Simulation}\label{sec:qpe}
Quantum Phase Estimation (QPE) extracts ground or excited state energies when the prepared initial state has a large overlap with the desired target eigenstate. 
A promising alternative to canonical QPE uses Hamiltonian simulation with a Hadamard test to measure the complex overlap between the initial state and the time-evolved states at several time steps. The resulting time-series is then post-processed, e.g.\ by quantum complex exponential least-squares (QCELS) \cite{ding23}, to extract the energies. Similarly to the canonical QPE, the performance of the Hamiltonian simulation with QCELS depends sensitively on the initial state, therefore it is of interest to prepare the initial state in the most resource efficient way. Formally, the QCELS method requires the measurement of
\begin{equation}
Z(t)=\bra{\psi}e^{-it \hat{H}}\ket{\psi},
\label{eqn:zed_t}
\end{equation}
where if $\ket{\psi} = \ket{\psi_{0}}$ is the exact ground state, (or an eigenstate), then it coincides with $e^{-it \bra{\psi_0}\hat{H}\ket{\psi_0}}$, from which the energy $E_0 = \bra{\psi_0}\hat{H}\ket{\psi_0}$ can be easily extracted. Typically, however $\ket{\psi}$ is not an eigenstate. Following the QCELS procedure we sample the time $t_{n}=n\tau$ with $n=0,\dots ,N-1$ and time step $\tau$ and obtain an estimate of $Z(t_n)$ via the Hadamard test. We prepare \begin{equation}
\ket{\Psi(t_{n})}=
\frac{1}{\sqrt{2}}\bigl(\ket{0}\otimes\ket{\psi}
                      +\ket{1}\otimes U(\tau)^{n}\ket{\psi}\bigr),
\end{equation} where $U(\tau)=e^{-i\tau \hat{H}}$, and the real and imaginary parts are estimated via the expectation values
\begin{equation}
\Re Z_{n}=\bra{\Psi(t_{n})}X\otimes I\ket{\Psi(t_{n})},
\end{equation} and
\begin{equation}
\Im Z_{n}=\bra{\Psi(t_{n})}Y\otimes I \ket{\Psi(t_{n})}.
\end{equation} The energy is then extracted from the time-series by maximizing the QCELS objective
\begin{equation}
f(E)=\left|\sum_{n=0}^{N-1} Z(t_{n})\,e^{in\tau E}\right|^{2}.
\label{eqn:qcles_func}
\end{equation} and the location of the maximum yields the energy estimation. The time step is chosen such that $\tau<\frac{2\pi}{E_{\text{max}} - E_{\text{min}}}$, where $E_{\text{max}}$  and $E_{\text{min}}$ are the largest and smallest eigenvalues of the simulated Hamiltonian. This choice ensures $\tau E\in(-\pi,\pi)$ for every eigenvalue $E$, therefore avoiding phase aliasing and enabling unambiguous extraction via Eq. \ref{eqn:qcles_func}. In practice, the spectrum can be centered by using $H-\frac{\operatorname{Tr}(H)}{2^{n_q}}$, so that the ground state energy corresponds to the smallest eigenphase.

In Sec. \ref{results} we demonstrate how multiconfigurational state preparation can increase the overlap with the ground state and therefore impacts the performance of the QCELS method.

\subsubsection{Quantum Self-Consistent Equation Of Motion}\label{sec:qsceom}

The calculation of excited states has attracted the interest of the quantum computing community as a first step towards the calculation of response properties. Quantum self-consistent equation of motion (Q-SCEOM) was suggested recently for the calculation of excitation energies \cite{asthana23} and response properties \cite{Kumar2023}. Q-SCEOM uses self-consistent excitation operators in order to satisfy the vacuum annihilation condition even when an approximate set of excitation operators is used \cite{asthana23, Mukherjee1985, Mukherjee1993}. As a result of this, Q-SCEOM yields accurate results in cases where plain qEOM \cite{qEOM2020} fails qualitatively (see recent works \cite{asthana23, Kumar2023, Motta2024, Asthana2025} for comparisons between Q-SCEOM, qEOM, and quantum subspace expansion \cite{mcclean17, mcclean20}). As shown by Asthana \textit{et al}. \cite{asthana23}, the use of self-consistent excitation operators simplifies the final EOM equations (\textit{cf} qEOM). The excitation energies are obtained by diagonalization of the $M$ matrix whose elements are given by (see Asthana \textit{et al.} \cite{asthana23} for a detailed derivation):

\begin{align}\label{eq:sceom_energy}
    M_{IJ} = \left<\psi_{\mathrm{HF}}|\hat{G}_I^\dagger U^\dagger(\vec{\theta}_{\mathrm{opt}})\hat{H}U(\vec{\theta}_{\mathrm{opt}})\hat{G}_J|\psi_{\mathrm{HF}}\right> - \delta_{ij}E_{\mathrm{gr}}
\end{align}
where $\hat{G}_I$ are excitation operators, $\hat{H}$ is the system's Hamiltonian, and $\left|\psi_{\mathrm{HF}}\right>$ is the Hartree-Fock (HF) state. $U(\vec{\theta}_{\mathrm{opt}})$ is a unitary ansatz with angles $\vec{\theta}_{\mathrm{opt}}$ optimized for the ground state.
A simplified form of the off-diagonal elements can be obtained \cite{asthana23}
\begin{align}
    \mathrm{Re}\left[M_{I,J}\right] = M_{I+J,I+J} - \frac{M_{I,I}}{2} - \frac{M_{J,J}}{2}
\end{align}
where 
\begin{align}\label{eq:sceom_off}
    &M_{I+J,I+J} = \\ \nonumber
    &\frac{1}{2}\left<\psi_{\mathrm{HF}}|\left(\hat{G}_I^\dagger + \hat{G}_J^\dagger\right) U^\dagger(\vec{\theta}_{\mathrm{opt}})\hat{H}U(\vec{\theta}_{\mathrm{opt}})\left(\hat{G}_I +\hat{G}_J\right)|\psi_{\mathrm{HF}}\right>
\end{align}

As shown in Eq.~\ref{eq:sceom_off}, the calculation of the off-diagonal elements of the $M$ matrix requires the construction of a multiconfigurational state, $\left|\psi_{I+J}\right> = \left|\left(\hat{G}_i +\hat{G}_j\right)|\psi_{\mathrm{HF}}\right>$ which is the most demanding step for the quantum computation part. 

Here we discuss the application of the  multiconfigurational state preparation methods presented in Sec. \ref{methods}, comparing both approaches, each being a building block of the Q-SCEOM implementation in InQuanto \cite{inquanto_web, inquanto_docs}.

\subsection{Computational Details}\label{sec:comp_details}

To demonstrate the use of multiconfigurational state preparation as a useful approach to prepare initial states for strongly correlated systems, we study the ground and excited states of twisted C$_2$H$_4$ in the minimal basis set STO-3G. The ground state corresponds to a singlet at torsion angles close to 0\textdegree\ or 180\textdegree, while at 90\textdegree\ the ground state is quasi-degenerate with the triplet having a slightly lower energy.   In terms of classical preprocessing, the PySCF code \cite{pyscf} is used to build the fermionic (second quantized) Hamiltonian and to run complete active space configuration interaction (CASCI) and unrestricted HF calculations. Active spaces are selected based on the ordering of mutual information \cite{boguslawski15} between orbital pairs (see Fig.~ \ref{app:orbs_plot} in the Appendix for the plots of the molecular orbitals (MOs) that were used in the active spaces). For all active spaces and all torsion angles, the natural orbitals are dominated by the carbon $2p$ shell, with small admixtures of H $1s$ that grow (but remain minority) as the torsion angle tends to 90\textdegree. For larger (8-qubit and 12-qubit) active spaces, higher orbitals that contribute to correlation in the ground state also contain mixtures of $2s$ and $2p$. Fermionic Hamiltonians are JW transformed to obtain $\hat{H}$ as a sum of Pauli strings, for each active space. For VQE optimizations, the L-BFGS-B method as implemented in the SciPy package \cite{scipy} was used. The multiconfigurational state preparation methods are implemented in the InQuanto software package \cite{inquanto_web, inquanto_medium, inquanto_docs}, and all quantum calculations are carried out using \mbox{InQuanto}. Quantum circuits are compiled using the architecture agnostic software compiler TKET \cite{tket20}.

\section{Results} \label{results}

\subsection{Ground State of C\texorpdfstring{$_2$}{2}H\texorpdfstring{$_4$}{4}}\label{sec:results_gs}

Here we demonstrate the use of multiconfigurational state circuits to calculate the ground state energy of twisted C$_2$H$_4$, using variational and non-variational techniques. A common bottleneck in these eigensolvers is the large overhead in measurements needed to accurately represent the required expectation values (typically by averaging over Pauli string measurements). This issue is exacerbated in strongly correlated systems in which a single reference (single ON configuration) state does not have a large overlap with the true ground state. A warm start initial state can be highly beneficial in this regard, particularly for quantum subspace methods, as the measurement overhead can be reduced, for example by requiring a smaller number of Hamiltonian moments or a smaller dimensionality of the subspace. 

Before discussing results for the ground state energies of specific active spaces, we mention a popular quantum computational method to approximate the ground state which is Unitary Coupled Cluster Singles and Doubles (UCCSD) \cite{anand22}, with excitation parameters $\theta_k$ optimized by VQE. This ansatz can be expressed as a series of unitary operators applied to a reference state (e.g. to the Hartree-Fock state $|\psi_{\mathrm{HF}}\rangle$)

\begin{equation}\label{eqn:ucc}
    |\psi_{\mathrm{UCC}}\rangle = e^{\sum_k \theta_k (\hat{T}_k - \hat{T}^\dagger_k)}|\psi_{\mathrm{HF}}\rangle \ ,
\end{equation}

\noindent where the fermionic excitation operators $\hat{T}_k$ consist of single, double, triple, ..., excitations, which excite 1, 2, 3, ..., fermions between occupied and virtual orbitals. In UCCSD, the excitations are truncated to singles and doubles, corresponding to each $\hat{T}_k$ operating over a maximum four spin oribtal indexes. (In practise, a Trotterized form of UCCSD is used as the variational ansatz, since individual excitations may not commute. Here we omit the Trotterization from Eq. \ref{eqn:ucc} for brevity).

\begin{figure*}[ht]
\centering
        \includegraphics[width=0.5\linewidth]{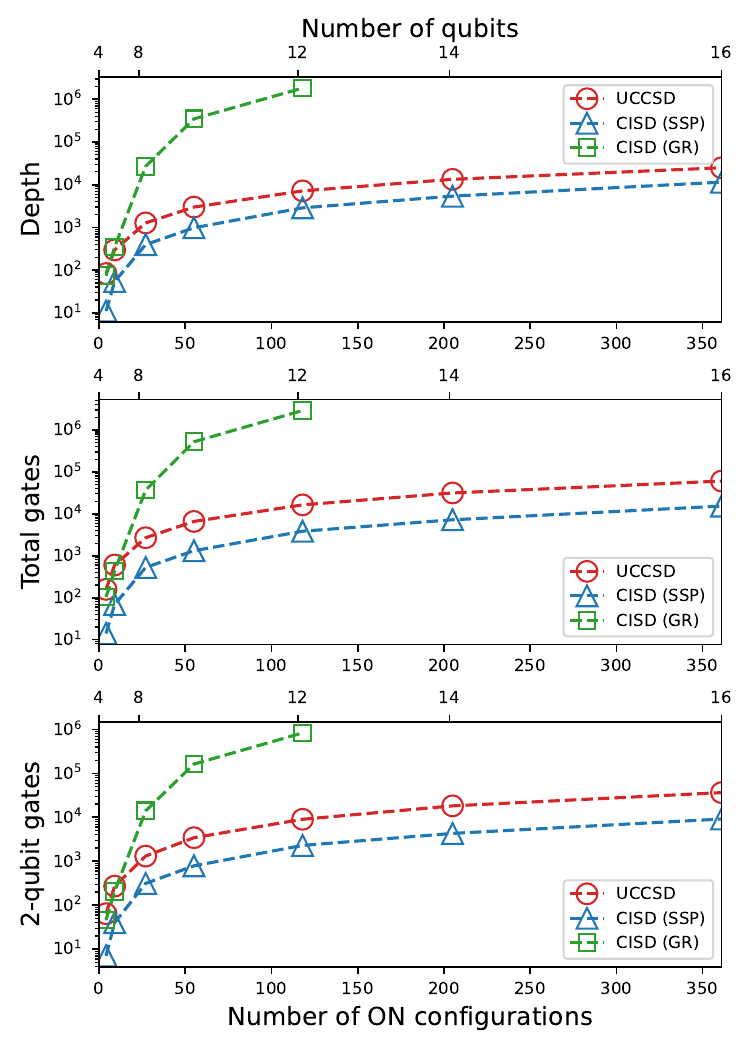}
        \caption{Circuit resources obtained for UCCSD and CISD, where the latter are prepared using the GR (Sec. \ref{sec:gr_method}) or the SSP \cite{gleinig21} (Sec. \ref{sec:ssp_method}) methods. Number of qubits $n_q \in \{4, 6, 8, 10, 12, 14, 16\}$. For a given $n_q$ (equal to number of spin orbitals), the number of electrons is $\frac{n_q}{2}$ if $\frac{n_q}{2}$ is even or $\frac{n_q - 1}{2}$ if $\frac{n_q}{2}$ is odd, for which the HF reference is a closed shell singlet configuration. For a given $n_q$ and number of electrons, the number of configurations corresponds to the number of single and double excitations plus the HF reference. Circuits are compiled to the standard gate set using the Qiskit \cite{qiskit} extension of TKET \cite{tket20}.}
        \label{fig:cisd_resources}
\end{figure*}

If one applies the second quantized fermionic operators $T_k$ directly to $|\psi_{\mathrm{HF}}\rangle$, a series of ON configurations can be generated which correspond to all possible single and double excitations relative to the HF state. The latter in fact correspond to the basis states of Configuration Interaction Singles and Doubles (CISD) \cite{pink_book_ch1}, and the CISD wavefunction is obtained once the coefficients of those basis states are found (typically by diagonalization of the Hamiltonian in the CISD basis). While UCCSD can be more accurate than CISD (due to size extensivity, for example), CISD contains a significant portion of electronic correlation and often yields reasonable approximations to the ground state. Due to the framework we developed for preparing multiconfigurational states, a variational circuit corresponding to the CI expansion can easily be generated once the ON configurations of the expansion are given. In Fig \ref{fig:cisd_resources}, we compare CISD circuit sizes prepared using the SSP and GR methods to UCCSD circuits (with all gate angles represented symbolically) for a range of active spaces. For each active space, the ON configurations are obtained by applying $\hat{T}_k$  to the closed shell singlet HF state. Therefore, the states prepared using these ON configurations retain the same spin symmetry as the ground states of C$_2$H$_4$ for torsion angles close to 0\textdegree \ or 180\textdegree. 

We observe that the SSP method produces circuits significantly smaller than the GR method over the range of active spaces chosen (4 - 16 qubits correspond to 2-8 active molecular orbitals). This is primarily due to the scaling in the number of external controls required for the GR method as the state becomes more complex with size, in addition to the decomposition of GR 1-body and 2-body rotations. The circuits produced by SSP are also consistently smaller than those produced by UCCSD for the same active space, implying the lower susceptibility to gate fidelity errors and qubit decoherence of the SSP CISD circuits, despite the generally lower accuracy of CISD compared to that of UCCSD.

\subsubsection{4-Qubit Active Space: VQE} \label{sec:res_4qvqe}

\begin{figure}[ht]  
\centering
\includegraphics[width=1.0\linewidth]{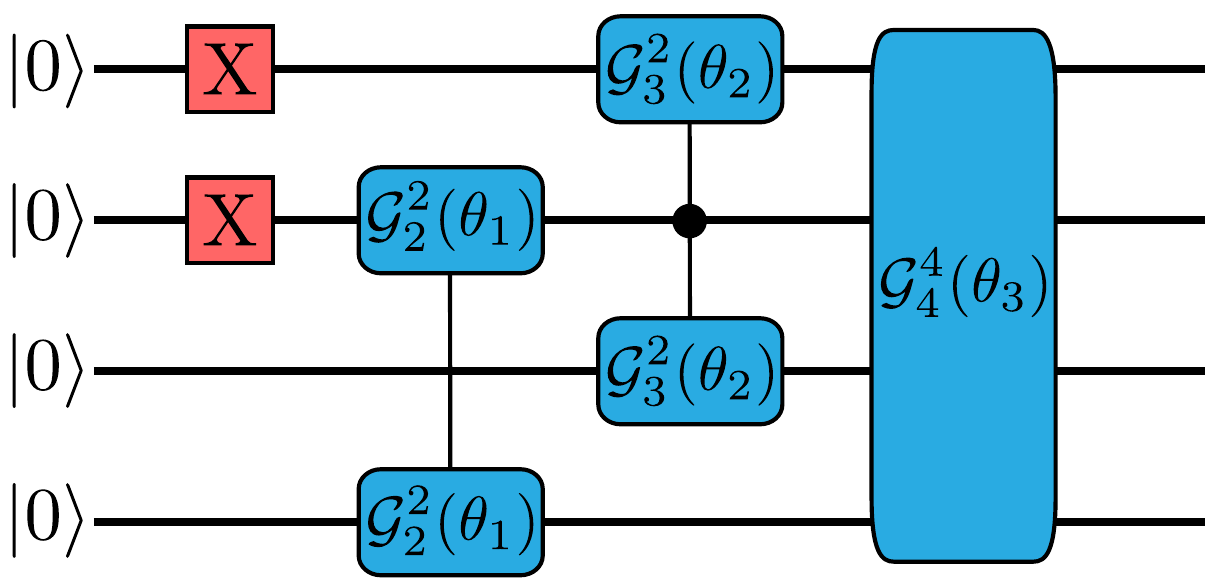}
        \caption{Circuit corresponding to state $c_1|1100\rangle + c_2|1001\rangle+c_3|0110\rangle+c_4|0011\rangle$ prepared using the GR method (see Sec. \ref{sec:gr_method}). Algorithm \ref{alg:gr} found that externally controlling $\mathcal{G}_3^2$ on the second qubit is required. Substituting the gate parameters for 90\textdegree\ torsion as an example and compiling to the H-series gate set \cite{h11}, this circuit can be represented using 61 PhasedX gates, 4 Rz gates, and 44 2-qubit ZZMax gates.}
        \label{fig:gr_4q}
\end{figure}

\begin{figure}[ht]  
\centering
\includegraphics[width=1.0\linewidth]{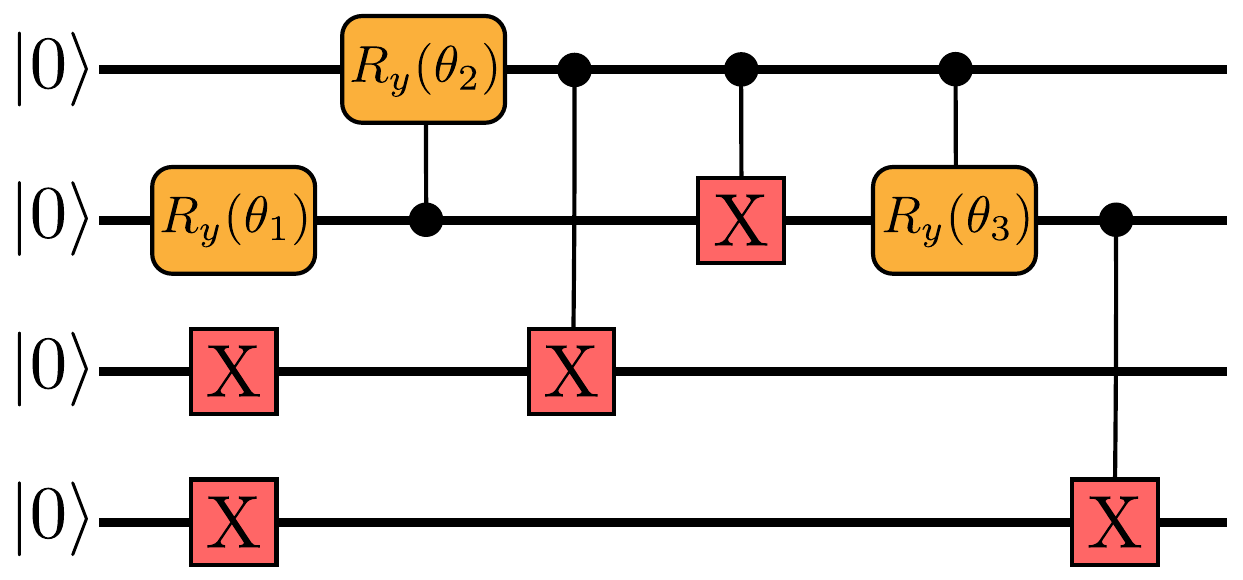}
        \caption{Circuit corresponding to state $c_1|1100\rangle + c_2|1001\rangle+c_3|0110\rangle+c_4|0011\rangle$ prepared using the SSP method \cite{gleinig21} (see Sec. \ref{sec:ssp_method}). Substituting the gate parameters for 90\textdegree\ torsion as an example and compiling to the H-series gate set \cite{h11}, this circuit can be represented using 10 PhasedX gates, 4 Rz gates, and 5 2-qubit ZZMax gates.}
        \label{fig:ssp_4q}
\end{figure}

\begin{figure*}[ht]
\centering
        \includegraphics[width=0.7\linewidth]{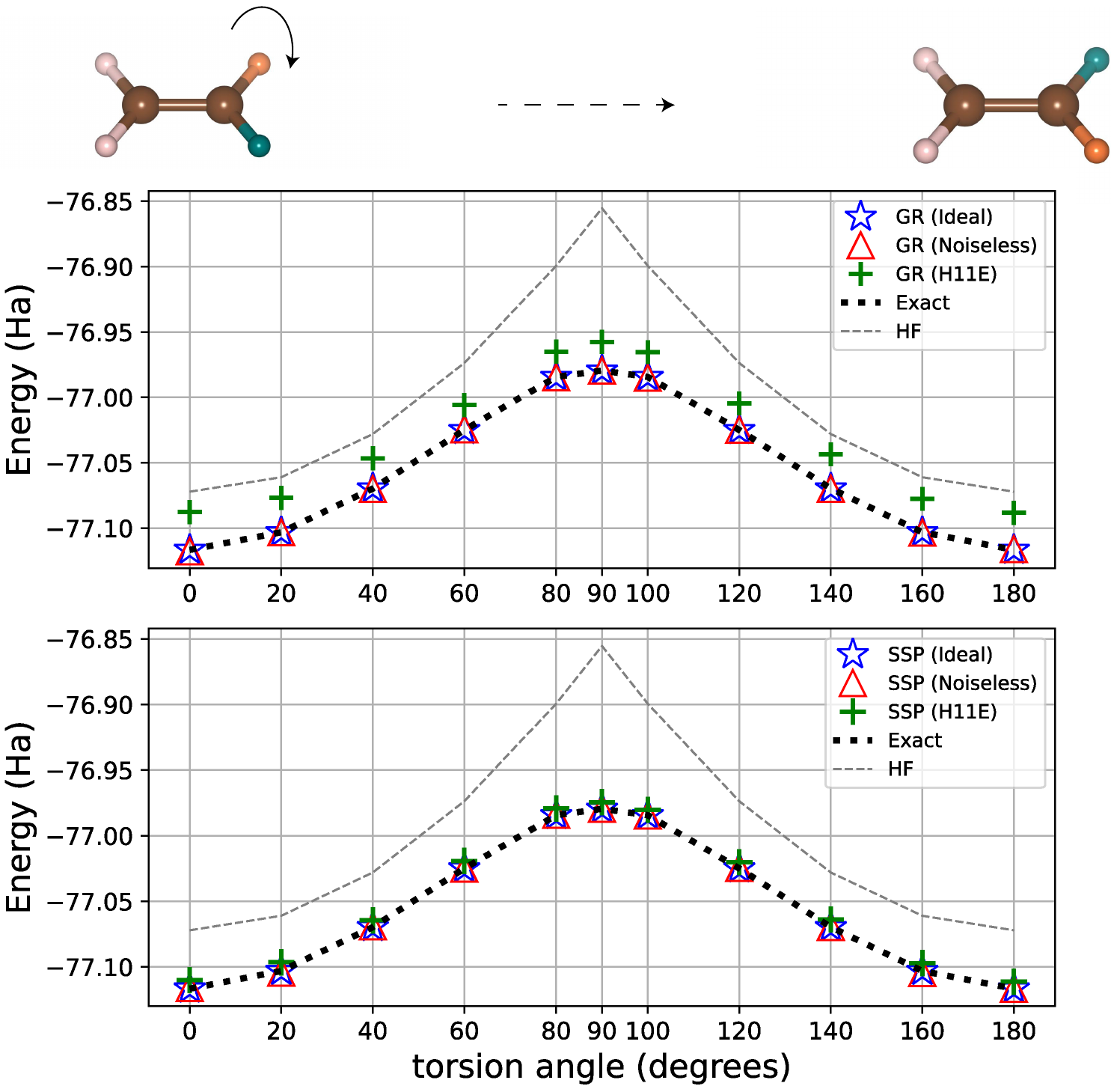}
        \caption{Energies obtained from VQE-optimized multiconfigurational states for the 4-qubit (2 electrons in 4 spin orbitals) active space of C$_2$H$_4$. Simulated measurements correspond to 10$^4$ shots per circuit, and the Hamiltonian consists of 14 Pauli operators. Top graph: GR method (see Sec. \ref{sec:gr_method}). Bottom graph: SSP method (see Sec. \ref{sec:ssp_method}). C$_2$H$_4$ structure at torsion angles 0\textdegree, 90\textdegree, and 180\textdegree \ shown above graphs. H11E corresponds to emulations of hardware experiments with a noise model calibrated to the H1 trapped ion device \cite{h11}.
        }
        \label{fig:vqe_2in4}
\end{figure*}

As a first demonstration of multiconfigurational state preparation methods, we consider ansatzes for VQE applied to twisted C$_2$H$_4$ in an active of space $n_\text{MO}=2$ molecular orbitals (MOs), corresponding to 4 qubits in the JW representation, populated by 2 electrons. Representing the ON configurations as $|q_1, q_2, ..., q_{n_q}\rangle$ where $n_q = 2n_\text{MO}$, the HF state can be written in a bit string representation as $|1100\rangle$. The ground state at equilibrium is a closed shell singlet, which transitions to a triplet state $\langle S^2 \rangle=2$ at 90\textdegree\ torsion angle. Hence, to capture the lowest energy manifold throughout the 180\textdegree\ rotation the following ON configurations are selected and included in the variational ansatz $|\psi(\vec{\theta})\rangle = \sum_d c_d|x_d\rangle$ for VQE optimization

\begin{equation}\label{eqn:vqe_4q}
    \begin{split}
    x_1 &= |1100\rangle, \\
    x_2 &= |1001\rangle, \\
    x_3 &= |0110\rangle, \\
    x_4 &= |0011\rangle \ .
    \end{split}
\end{equation}

\noindent The relation between variational parameters $(\vec{\theta})$ and coefficients $c_d$ are described as follows. 

For the GR method, as described in Sec. \ref{sec:gr_method} each GR linearly combines the reference ($x_1$) with another ON configuration in the input set. 3 GRs are required to linearly combine the configuration pairs $(x_1, x_2), (x_1, x_3), (x_1, x_4) \rightarrow \mathcal{G}_2^2(\theta_1), \mathcal{G}_3^2(\theta_2), \mathcal{G}_4^4(\theta_3)$. In this case, externally controlling $ \mathcal{G}_3^2(\theta_2)$ on the second qubit ($q_2$) is required to maintain the desired state vector (see Fig. \ref{fig:gr_4q}): since $x_2 = |1001\rangle$ is ``rotatable'' by $\mathcal{G}_3^2(\theta_2)$, then the external control is placed so that $\mathcal{G}_3^2(\theta_2)$ only operates on basis states in which $q_2=1$ (and hence has no effect on $|1001\rangle$). The corresponding pairs of coefficients (elements of the $\mathcal{G}_e$ matrices related to rotation angle, see Eq. \ref{eqn:2qGR}) are obtained through recursive normalization of the $c_d$ state vector coefficients \cite{arrazola22}, which here become $(\alpha_1=\cos\theta_1, c_2'=\sin\theta_1), (\alpha_2=\cos\theta_2, c_3'=\sin\theta_2), (\alpha_3=\cos\theta_3, c_4'=\sin\theta_3)$ where $\alpha_d = \sqrt{1-c_{d+1}'}$ and $c_{d+1}'=c_{d+1}/\prod_{b=0}^{d-1}\alpha_b$ with $\alpha_0 = 1$\footnote{This is consistent with $|\psi\rangle = \alpha|x_1\rangle + \sum_{d=2}^{D}c_d|x_d\rangle$ where $\alpha = \prod_{a=1}^{D-1}\alpha_a$. The sequence of GRs applied to the circuit  returns the desired state vector through this recursive mapping of the normalized coefficients, as discussed in Arrazola \textit{et al.} \cite{arrazola22}.}. For the SSP method, coefficients are related to gate angles as described in sec. \ref{sec:ssp_method} and exemplified in Gleinig and Hoefler's paper \cite{gleinig21}.

Figs. \ref{fig:gr_4q} and \ref{fig:ssp_4q} show the VQE ansatz circuits of 4 ON configurations (Eq. \ref{eqn:vqe_4q}), built using the SSP and GR methods. Depending on the optimized VQE parameters $\vec{\theta} = \{\theta_1, \theta_2, \theta_3\}$, this ansatz spans both singlet and triplet eigenstates (with spin number restricted to $s_z = 0$). For example, at torsion angle = 0\textdegree, optimization of $\vec{\theta}$ results in (up to global phase) a doubly excited closed shell singlet with negligible contributions from singly excited configurations \mbox{$|\psi(\vec{\theta})\rangle \approx 0.96814|1100\rangle - 0.25045|0011\rangle$} (with $|c_2|, |c_3| <  2\times10^{-6}$). Whereas for torsion angle = 90\textdegree, with $s_z=0$ the ground state is dominated by a superposition of open shell configurations, with small but non-negligible contributions from the closed shell configurations. $|\psi(\vec{\theta})\rangle \approx -0.00009|1100\rangle + 0.70710|1001\rangle+0.70712|0110\rangle + 0.00007|0011\rangle$ (as all 4 ON configurations have non-negligible weight, this state vector is used as the example for circuit resources reported in the captions of Figs. \ref{fig:gr_4q} and \ref{fig:ssp_4q}). We note that following the optimization of all VQE parameters for torsion angles between 0\textdegree\ and 180\textdegree, ideal energies $\langle\psi(\vec{\theta})|\hat{H}|\psi(\vec{\theta})\rangle$ match the lowest eigenvalues obtained from exact diagonalization of $\hat{H}$ in the 2-particle sector (see Fig. \ref{fig:vqe_2in4}) (and in the absence of device or measurement noise, ideal energies obtained from the GR and SSP methods are identical). Hence, the multiconfigurational state ansatz reproduces the ground state manifold for all torsion angles after VQE optimization.

Emulations of quantum hardware experiments use a noise model calibrated to the H1 trapped ion quantum computer \cite{h11}, labeled as ``H11E''. As shown in Fig. \ref{fig:vqe_2in4}, quantum computations of the expectation value are subject to greater amounts of noise for the GR method due to the larger circuits (compare circuit resources reported in Figs. \ref{fig:gr_4q} and \ref{fig:ssp_4q}), related to the decompositions of particle-conserving GRs \cite{anselmetti21, arrazola22} in addition to the external control (for $\mathcal{G}_3^2$) required for this case. Despite the negligible coefficients of singly excited configurations for some torsion angles $\neq$ 90\textdegree, we retain all 4 ON configurations in the circuit state vector for all torsion angles for ease of comparison. However, we note that shorter circuits (which represent only 2 ON configurations) for some torsion angles (at this active space) will likely lead to similar ideal energies. 

Despite the smaller circuits produced by SSP for the same state, and hence its lower susceptibility to device errors, the GR method does exhibit a conceptual advantage over SSP, which can be particularly useful when interpreting a variational state in a chemical context: when all gate angles of a circuit produced by the GR method are 0, the state vector falls back to the first ON configuration of the input set ($x_1$ in $(x_1, ..., x_D)$). If $x_1$ is chosen to be the HF state, then the GR unitaries added to this circuit can be readily interpreted as excitations on top of the HF reference. This is not necessarily the case for the SSP method, whose $\vec{\theta}=0$ state is non-trivial to predict as it depends on the distribution of binary values throughout the set of input ON configuration bit strings \cite{gleinig21}. Hence, the reference state for chemical excitations is not necessarily accessible as the $\vec{\theta}=0$ state of SSP method. This issue can be exacerbated in variational optimizations of the gate angles, which can terminate in local minima at small gate angles: for the SSP method this local minimum state can be unpredictable, whereas for the GR method it is likely close or equal to the HF state. For the small VQE applications presented here, this issue can be bypassed by random initialization of SSP parameters (avoiding $\vec{\theta}=0$ initialization). However, difficulties may arise in larger systems with more complicated parameter landscapes. 

\subsubsection{8-Qubit Active Space: Hamiltonian Moments} \label{sec:res_8qqcm}

\begin{figure*}[ht]
\includegraphics[width=1.0\linewidth]{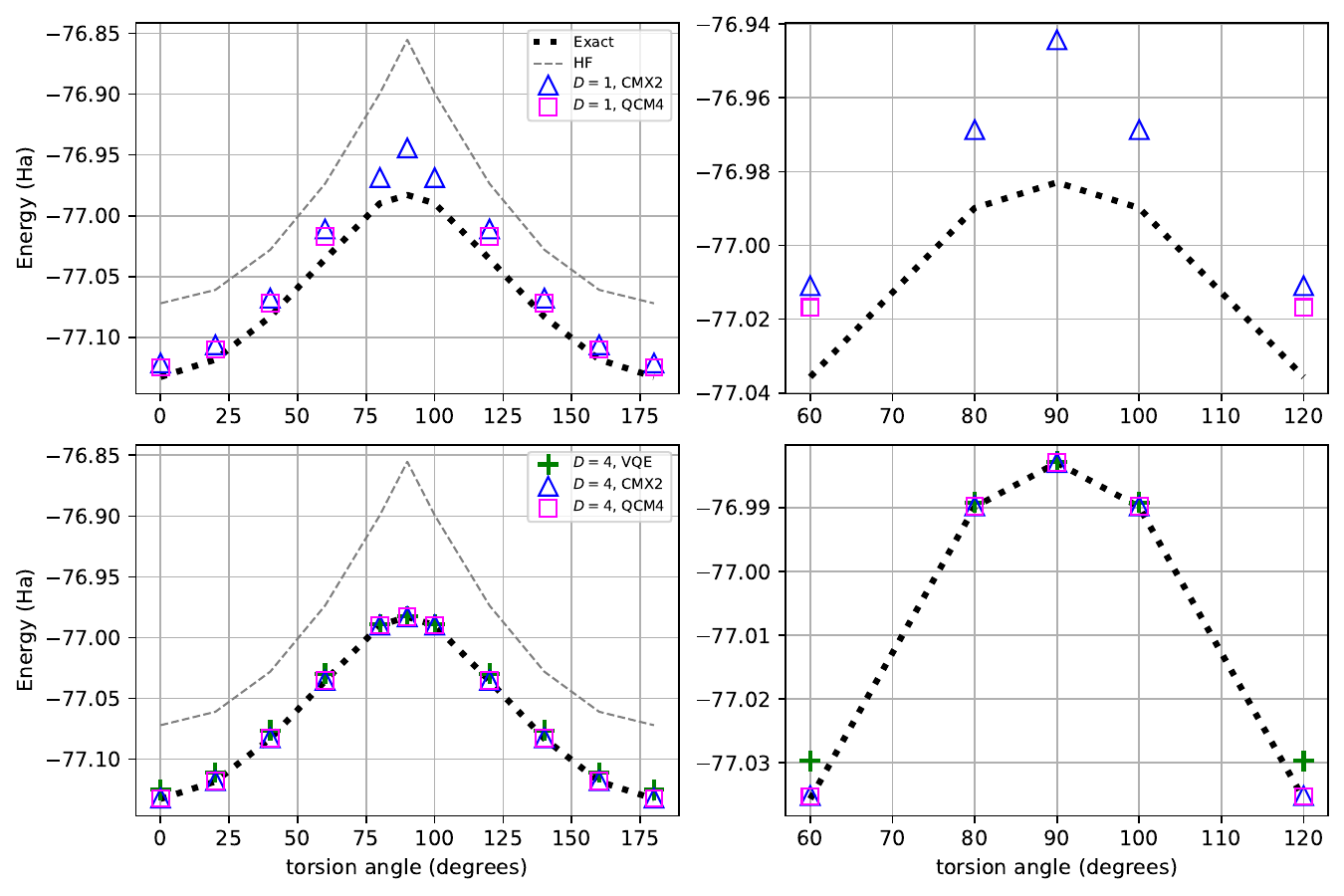}
        \caption{Ideal QCM4 and CMX2 energies calculated using a single configuration HF (top, $D=1$) and  VQE-optimized multiconfigurational input states (bottom, $D=4$) for the 8-qubit (4 electrons in 8 spin orbitals) active space of C$_2$H$_4$. VQE energies correspond to expectation values of $\hat{H}$ taken with respect to the optimized multiconfigurational state. For ideal simulations, externally controlled GRs (Sec. \ref{sec:gr_method}) and the SSP method (Sec. \ref{sec:ssp_method}) yield identical results. 
        }
        \label{fig:qcm4_cmx2_4in8}
\end{figure*}

For the 8-qubit active spaces, ON configurations are selected from the largest weight determinants observed in CASCI calculations. For torsional angles of 0\textdegree\ to 60\textdegree\ (and 120\textdegree\ to 180\textdegree), the ground state is dominated by the HF determinant ($|11110000\rangle$), consisting of mostly $2p$ orbitals in all MOs, with small mixtures of H $1s$ in the first ($q_1, q_2$) and forth ($q_7, q_8$) MOs. A paired double excited ON configuration ($|11001100\rangle$) is the next largest, followed by spin-paired open shell configurations ($|10011001\rangle, |01100110\rangle$). For torsion angle = 80\textdegree, the first two largest determinants are similar, while the open shell configurations instead have one MO (the third) fully occupied ($|10011100\rangle, |01101100\rangle$). At torsion angle = 90\textdegree, the ground state is a triplet, which is represented in its $s_z=0$ component using 4 ON configurations, each hosting open shell MOs ($|11100100\rangle, |11011000\rangle, |10110100\rangle, |01111000\rangle$). 8-qubit variational ansatzes of 4 ON configurations are then prepared for all torsion angles, which are optimized using ideal VQE. The resulting ON configurations and their coefficients are shown in Tab. \ref{tab:8qubit_states_gates}.

\begin{table*}
\small
\caption{State vectors and corresponding circuit resources for all torsion angles studied in the 8-qubit active space of C$_2$H$_4$. Note the symmetry around the 90\textdegree\ torsion angle (hence the circuits for 0\textdegree\ and 180\textdegree\ are equivalent, etc.). Circuit resources represent the GR method (SSP method in parenthesis). Circuits are compiled to the H-series gate set \cite{h11}.}
    \label{tab:8qubit_states_gates}
  \begin{tabular*}{\textwidth}
  {@{\extracolsep{\fill}}cccccccc}
    \hline
    Torsion angle (\textdegree) & $c_1|x_1\rangle$ & $c_2|x_2\rangle$ & $c_3|x_3\rangle$ & $c_4|x_4\rangle$ & PhasedX & Rz & ZZMax \\
    \hline
    0/180 & $0.9690|11110000\rangle$ & $-0.2345|11001100\rangle$ & \ \ $0.0546|10011001\rangle$ & $0.0547|01100110\rangle$ & 174 (22) & 7 (6) & 128 (17) \\
    20/160 & $0.9683|11110000\rangle$ & $-0.2380|11001100\rangle$ & \ \ $0.0533|10011001\rangle$ & $0.0534|01100110\rangle$ & 174 (22) & 7 (6) & 128 (17) \\
    40/140 & $0.9617|11110000\rangle$ & $-0.2648|11001100\rangle$ & \ \ $0.0503|10011001\rangle$ & $0.0503|01100110\rangle$ & 174 (22) & 7 (6) & 128 (17) \\
    60/120 & $0.9354|11110000\rangle$ & $-0.3481|11001100\rangle$ & \ \ $0.0441|10011001\rangle$ & $0.0441|01100110\rangle$ & 174 (22) & 7 (6) & 128 (17) \\
    80/100 & $0.8281|11110000\rangle$ & $-0.5522|11001100\rangle$ & $-0.0681|10011100\rangle$ & $0.0681|01101100\rangle$ & 66 (18) & 4 (3) & 40 (13) \\
    90 & $0.7044|11100100\rangle$ & \ \ $0.7044|11011000\rangle$ & \ \ $0.0615|10110100\rangle$ & $0.0615|01111000\rangle$ & 52 (16) & 4 (3) & 32 (11)  \\
    \hline
  \end{tabular*}
\end{table*}

The multiconfigurational states from Tab. \ref{tab:8qubit_states_gates} are then used as initial states for the QCM4 and CMX2 methods, with energies obtained from expectation values of Hamiltonian moments. Ideal results (noiseless and infinite shot limit) are shown in Fig. \ref{fig:qcm4_cmx2_4in8}. We first note that for torsion angles of 0\textdegree\ to 60\textdegree\ and 120\textdegree\ - 180\textdegree\ (representing relatively weak electronic correlation), the ideal energy values are reasonably accurate for the $D=1$ HF input state, and highly accurate for the $D=4$ multiconfigurational states. 

At torsion angles near the transition point (80\textdegree, 90\textdegree, 100\textdegree, strongly correlated as the singlet and triplet eigenstates become quasi-degenerate) we find that the $D=1$ closed shell HF input state results in numerically very small cumulants $\mathfrak{c}_n$, leading to the term $3\mathfrak{c}_3^2 - 2\mathfrak{c}_2\mathfrak{c}_4$ approaching 0 or negative and an ill-defined QCM4 formula when taking the square root (see Eq. \ref{eqn:qcm4}). Noting that the original derivation assumed the condition $3\mathfrak{c}_3^2 - 2\mathfrak{c}_2\mathfrak{c}_4 > 0$ \cite{hollenberg94}, we omit the QCM4 energies from these angles for $D=1$ (also considering that the closed shell HF state insufficiently represents the strongly correlated ground state, leading to a poor description of the correlation represented by the moment expectation values). However, for the $D=4$ multiconfigurational states, both the QCM4 and CMX2 calculations recover the exact lowest energies for all torsion angles. This shows the benefit of multiconfigurational input states for methods involving connected moments, as not only does the QCM4 formula avoid the issue of negative $3\mathfrak{c}_3^2 - 2\mathfrak{c}_2\mathfrak{c}_4$ (we assume due to a higher quality input state), but the lower order theory (CMX2) can obtain a similar accuracy, hence necessitating a lower order of moments and ultimately less Pauli strings to measure. The latter point is further emphasized by considering that the VQE energies at torsion angle near 90\textdegree\ are $<1$mHa above the exact values; at these geometries the expectation of $\hat{H}$ with respect to the multiconfigurational state already captures most of the electronic correlation, obviating the need for higher order moments with a small overhead in circuit depth (see circuit resources reported in Tab. \ref{tab:8qubit_states_gates}).

\subsubsection{12-Qubit Active Space: Quantum Phase Estimation and Hamiltonian Simulation}\label{sec:qpe_res}

\begin{figure*}[ht]  \includegraphics[width=0.9\linewidth]{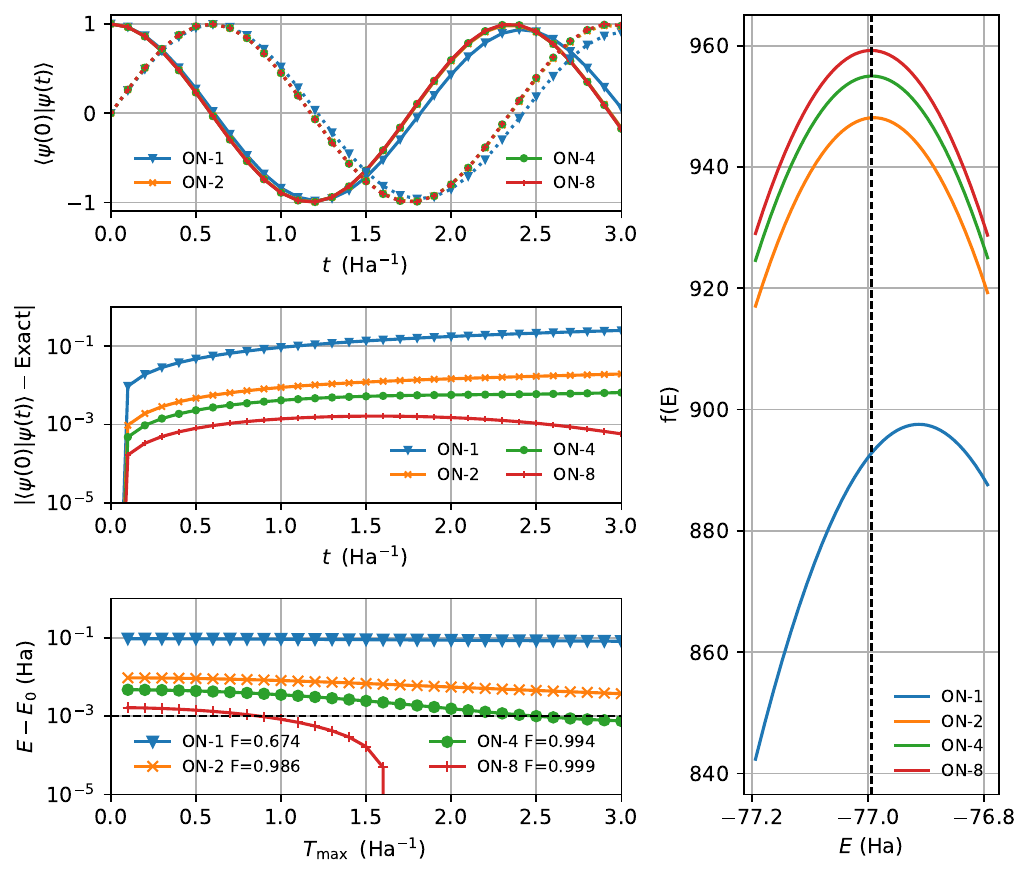}
        \caption{
        Top left panel shows $\bra{\psi(0)}\ket{\psi(t)}$ real (solid) and imaginary parts (dotted) for different initial states, for $\hat{H}$ representing the 12-qubit (6 electrons in 12 spin orbitals) active space for torsion angle = 80\textdegree. Middle left panel shows $|\bra{\psi(0)}\ket{\psi(t)} - \text{Exact}|$ for different initial states, exact overlap is calculated with ground state as initial state. Bottom left panel shows the QCELS energies for different initial states relative to the exact ground state energy. The x-axis is the time period for QCELS, demonstrating that the energy computed for that period requires $T_{\text{max}}$ simulation time. As the period increases, the energy approaches target accuracy (dashed horizontal line) with different rates for different initial state. Legend shows the fidelity between the initial state and the exact ground state. Right panel shows QCELS phase function for the largest $T_{\text{max}}$ period with the exact ground state energy marked by a dashed vertical line.
        }        \label{fig:qpe_80_qcels}
\end{figure*}

To quantitatively examine how the state preparation impacts the ground state estimation of QCELS, we performed exact state vector time evolutions for the $12$-qubit $(6e,6o)$ active space of C$_2$H$_4$ at a torsion angle of $80^{\circ}$. At this geometry the Hartree-Fock configuration, which has the largest weight in the exact ground state $\ket{\psi_0}$, has fidelity  only $| \braket{\psi_0}{\text{HF}} |^2=0.674$, which is below the $0.71$ threshold recommended in Ref. \cite{ding23} for reliable QCELS phase extraction. Moreover, it is expected that the required run time, therefore the circuit depth of the time evolution operator, becomes shorter as the initial state approaches the exact ground state \cite{ding23}. Therefore we performed simulations with initial states containing increasing number of ON configurations. All together four initial states were used in the simulations, denoted as
$\ket{\phi_{\text{ON-1}}}=\ket{\text{HF}},\,\,\,
\ket{\phi_{\text{ON-2}}},\,\,\,
\ket{\phi_{\text{ON-4}}},$ and $\ket{\phi_{\text{ON-8}}},$
where the last three are derived from CASCI and correspond to the $2$, $4$, and $8$ largest weight ON configurations, respectively. We also note the fidelities of the multiconfigurational states: $| \braket{\psi_0}{\phi_{\text{ON-2}}} |^2=0.986$, and $| \braket{\psi_0}{\phi_{\text{ON-4}}} |^2=0.994$, $| \braket{\psi_0}{\phi_{\text{ON-8}}} |^2=0.999$.

Figure~\ref{fig:qpe_80_qcels} summarizes the simulations.  
The top panel shows the complex time series $Z(t_n)$ (real and imaginary parts) for each initial state. All curves except the HF case lie almost on top of each other. The middle panel plots the relative error with respect to the exact value, $e^{-it_n\langle\psi_0|H|\psi_0\rangle}$. We chose $\tau = 0.1\ \text{Ha}^{-1}$ and $N=31$, giving a total evolution time $T_{\max}=(N-1)\tau$, long enough to capture at least one full period of the exact $Z(t)$. The right panel displays the QCELS objective, whose maximum was located with the BFGS method. Finally, the bottom-left panel shows the energy errors obtained by repeating the simulations with smaller $T_{\max}$ while keeping $N=31$, effectively choosing a smaller time step. The energies derived from the initial states $\ket{\phi_{\text{ON-1}}}$ and $\ket{\phi_{\text{ON-2}}}$ do not reach the typical target precision $\varepsilon = 1\text{mHa}$ within the time window, even at the largest $T_{\max}$, whereas those from $\ket{\phi_{\text{ON-4}}}$ and especially $\ket{\phi_{\text{ON-8}}}$ do, and importantly, the latter hits the target precision at approximately half the evolution time.

Because the depth of the circuit representation of $e^{-itH}$ scales linearly with $t$, compact, high fidelity state representation can save significant amount of gate operations by reducing the evolution time. 
Preparing $\ket{\phi_{\text{ON-4}}}$ via the SSP requires 13 two-qubit gates, while $\ket{\phi_{\text{ON-8}}}$ needs 40. Although the state preparation thus costs 27 additional 2-qubit gates, the shorter evolution time is expected to result a net saving well beyond the 27 gates.

\subsection{Excited States of C\texorpdfstring{$_2$}{2}H\texorpdfstring{$_4$}{4}}\label{sec:results_sceom}

\begin{figure*}[ht]
    \begin{minipage}[b]{\textwidth}
    \centering
        \includegraphics[width=0.8\linewidth]
        {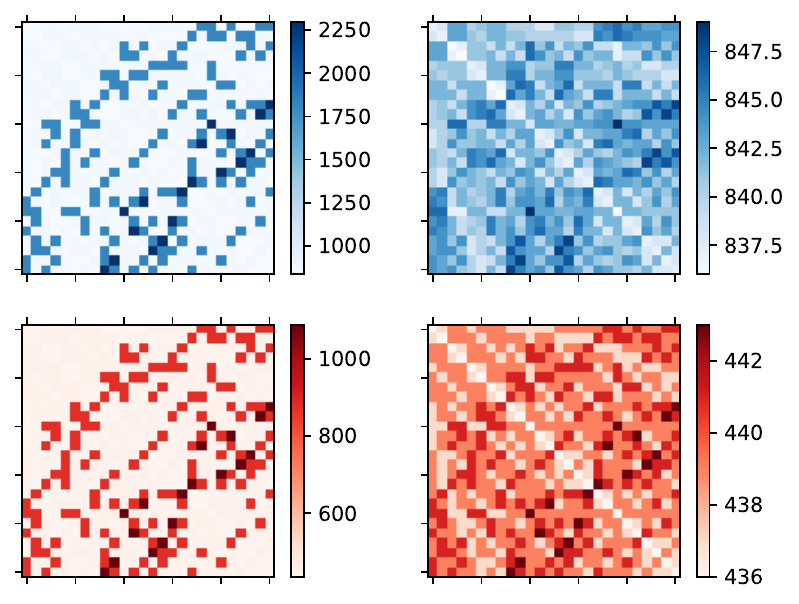}
        \caption{\small{\textbf{Top)} Total number of gates and \textbf{Bottom)} number of 2-qubit gates required for the construction states needed for the evaluation of the $M$ matrix ($U(\vec{\theta}_{\mathrm{opt}})\hat{G}_J\left|\psi_{\mathrm{HF}}\right>$ for the diagonal elements and $U(\vec{\theta}_{\mathrm{opt}})\left(\hat{G}_I +\hat{G}_J\right)\left|\psi_{\mathrm{HF}}\right>$ for the off-diagonal elements) for C$_2$H$_4$ with 8-qubit active space and 90\textdegree angle. The \textbf{Left)} GR, and \textbf{Right)} SSP methods were used for the off-diagonal terms. The circuits were compiled with the H-series emulator.}}
        \label{fig:gates_mats}
    \end{minipage}
\end{figure*}

\begin{figure*}[ht]
\centering
\includegraphics[width=0.75\linewidth]{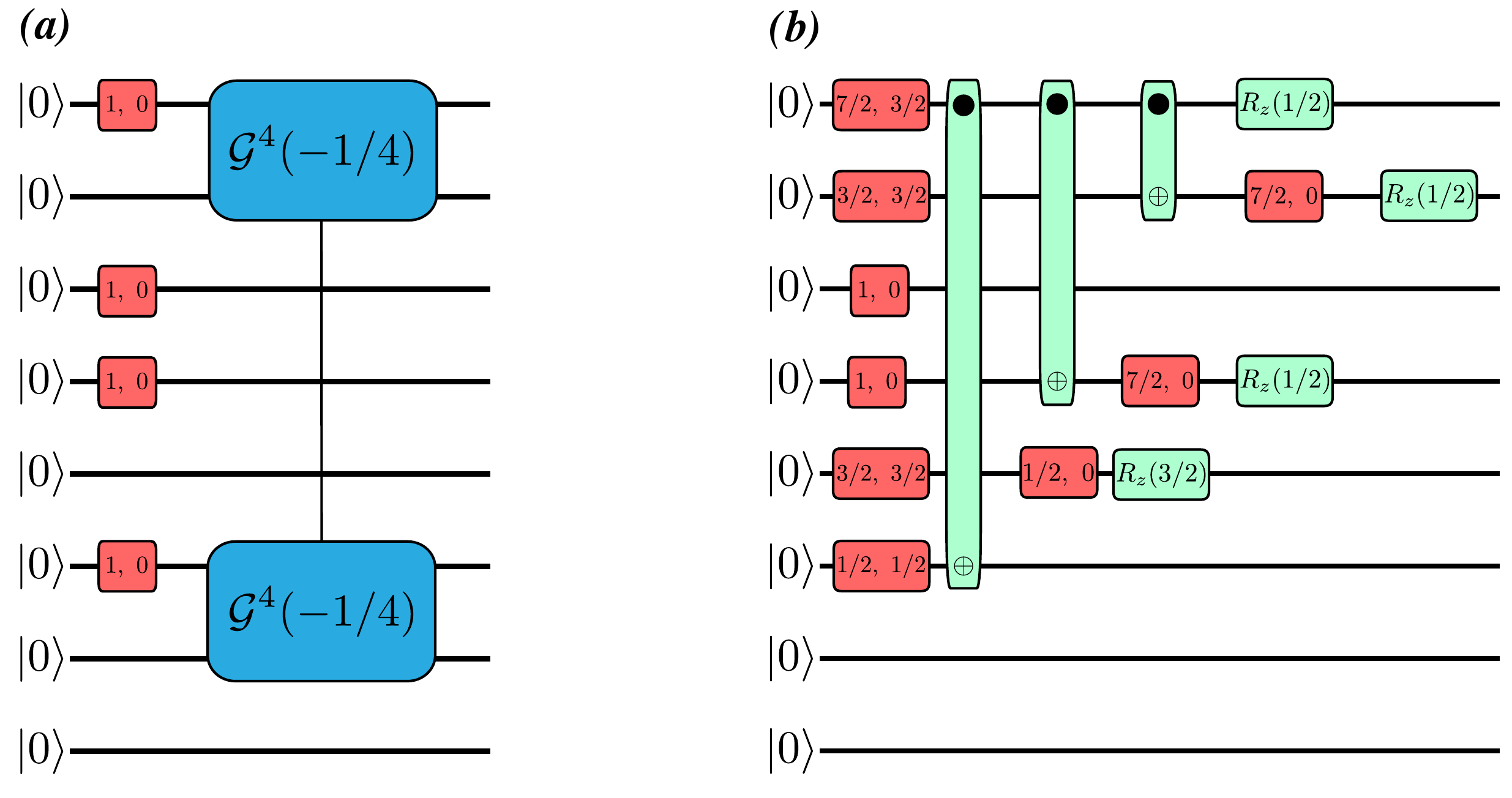}\hfill
\caption{\small{Circuits to construct the multiconfigurational state $\left|\left(\hat{G}_2 +\hat{G}_0\right)|\psi_{\mathrm{HF}}\right>$ for the $M_{20}$ element. Green vertical bars represent the ZZMax 2-qubit entangling gate, while red horizontal bars represent PhasedX($\alpha, \beta) = R_z(\beta)R_x(\alpha)R_z(-\beta)$ 1-qubit rotations \cite{h11}. The circuits represent (up to global phase) the state $\frac{1}{\sqrt{2}} \left( \left|10110100\right> - \left|01111000\right> \right)$. All gate angles are in units of $\pi$. (\fontfamily{ptm}\textit{\textbf{a}}) GR method, for which the $\mathcal{G}^4$ rotation decomposes into 14 ZZMax gates (see bottom panel of Fig. \ref{fig:gr2_gr4} for the representation of $\mathcal{G}^4$ in the H-series gate set). (\fontfamily{ptm}\textit{\textbf{b}}) SSP method, containing only 3 ZZMax gates.}}
\label{fig:sceom_circs}
\end{figure*}

\begin{figure*}[ht]
\centering
    \includegraphics[width=0.33\linewidth]{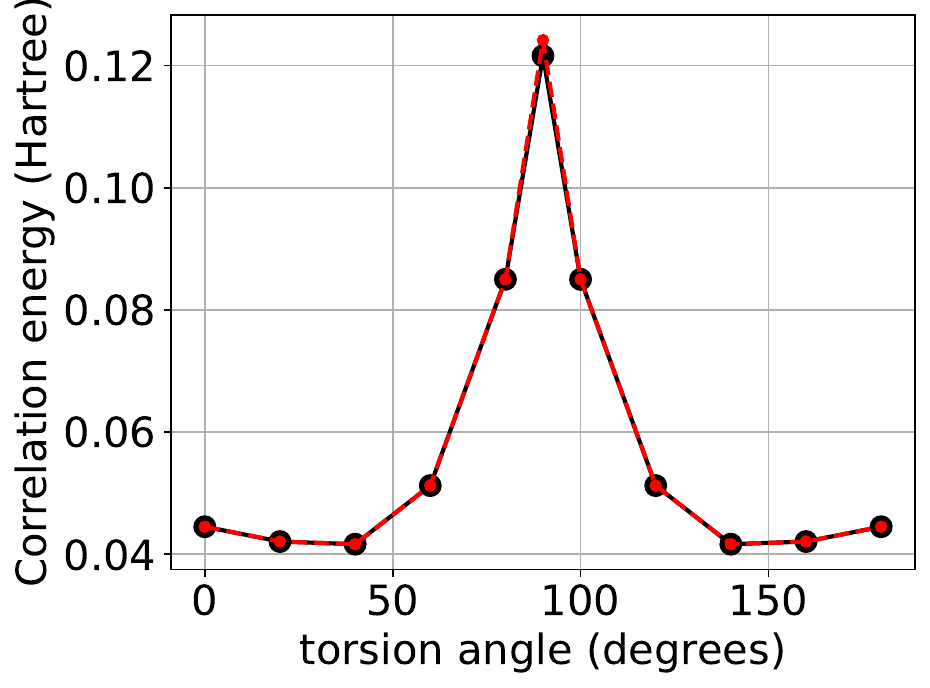}\hfill
    \includegraphics[width=0.33\linewidth]{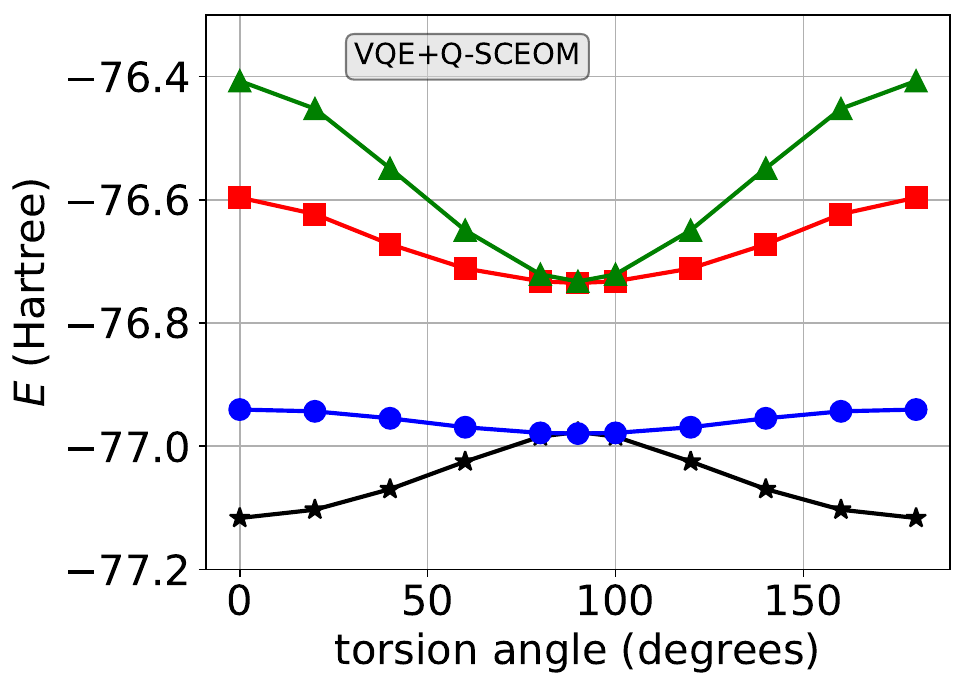}\hfill
    \includegraphics[width=0.33\linewidth]{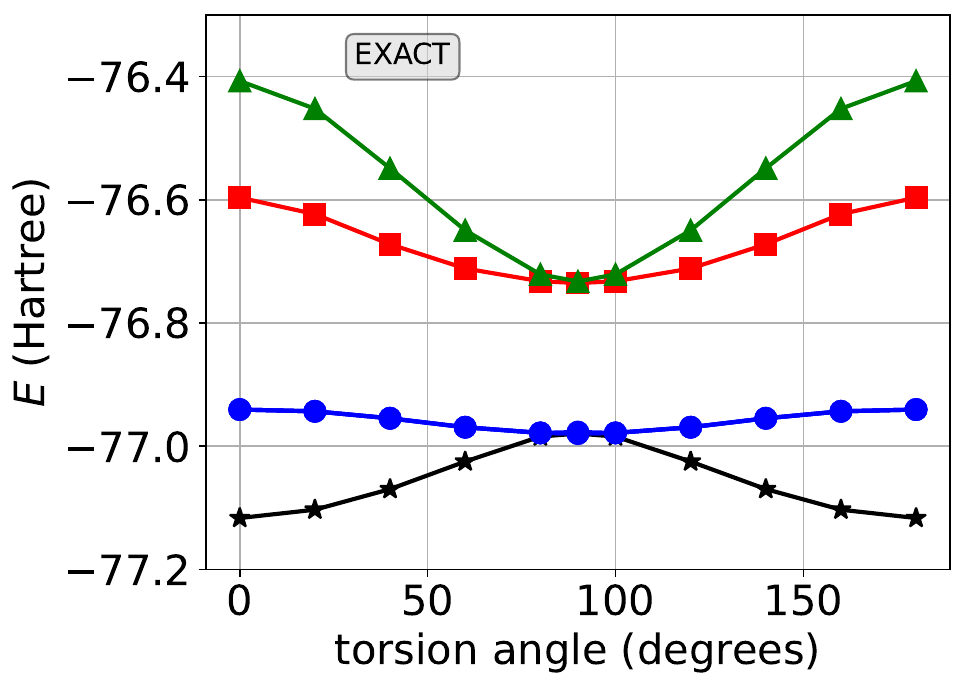}\hfill
    \includegraphics[width=0.33\linewidth]{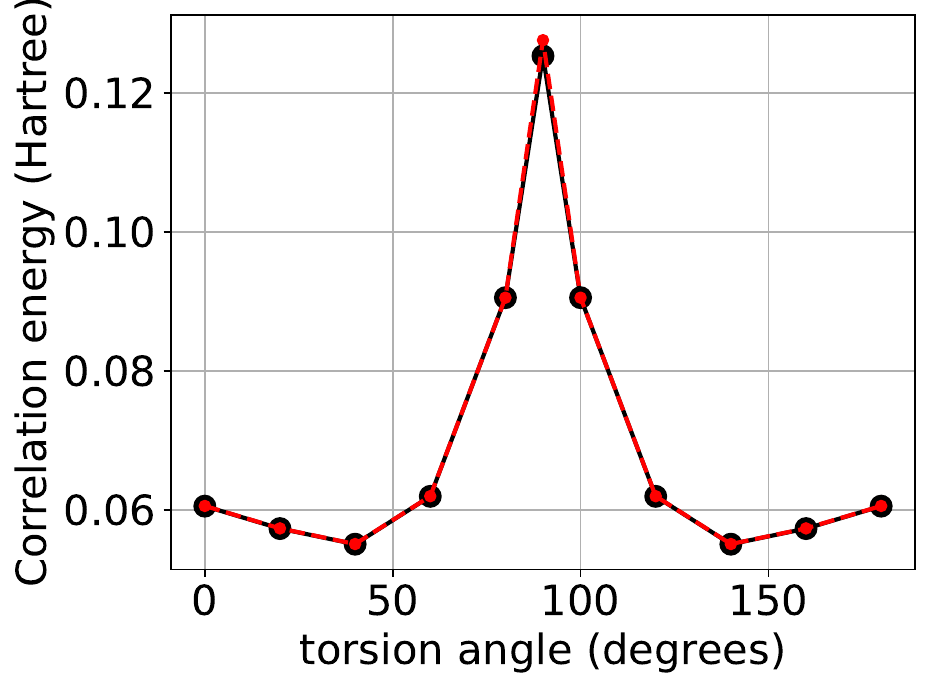}\hfill
    \includegraphics[width=0.33\linewidth]{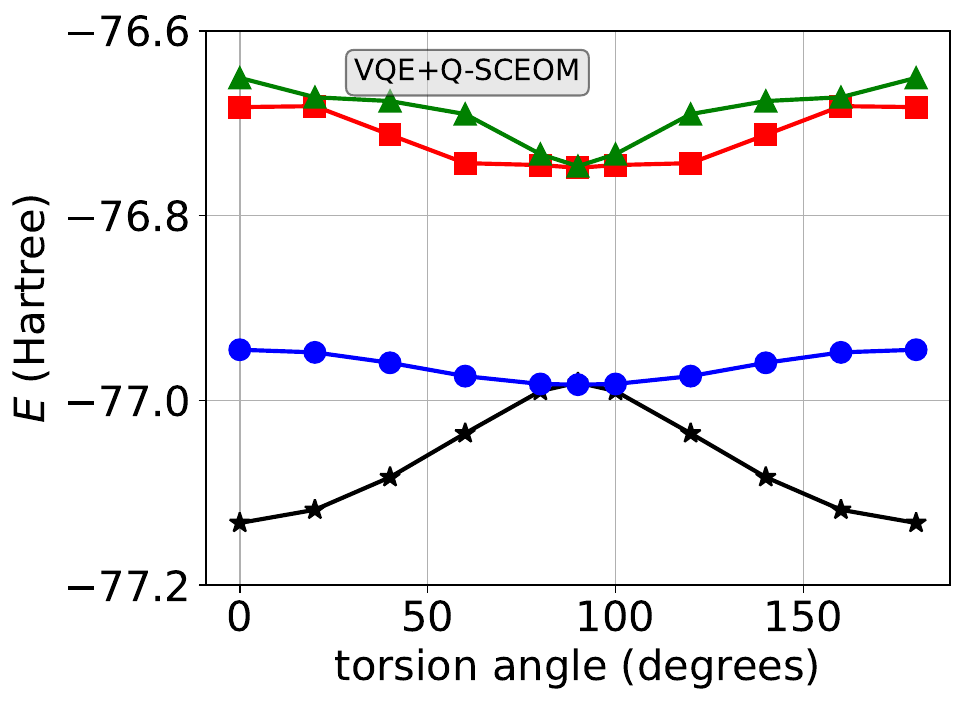}\hfill
    \includegraphics[width=0.33\linewidth]{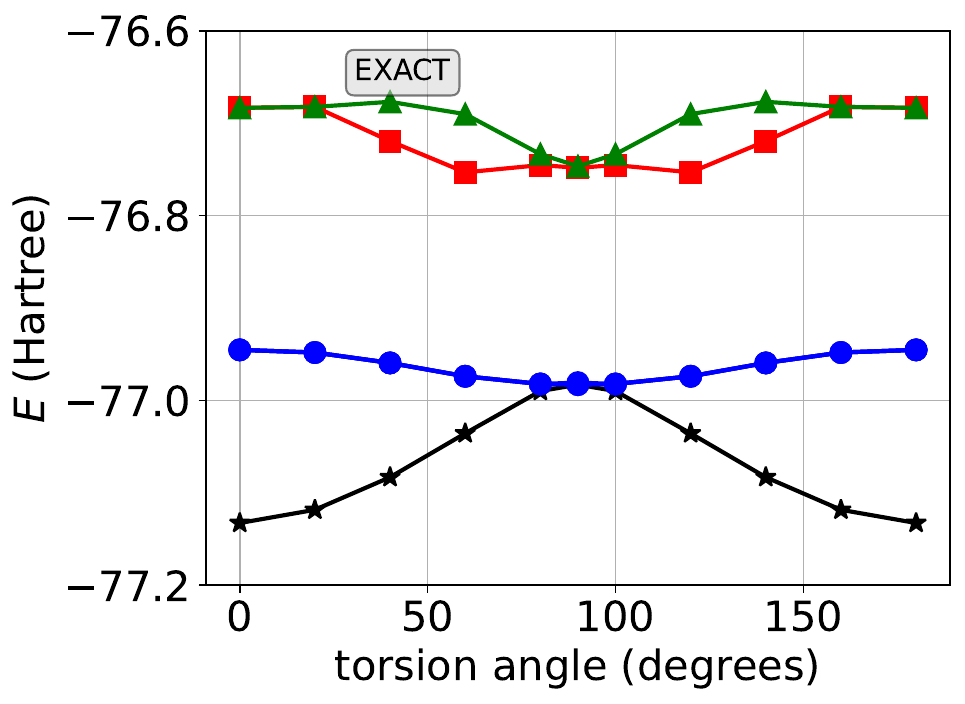}\hfill
\caption{\small{Noiseless calculations of ground and excited states for various torsion angles of 4-qubit and 8-qubit active spaces of C$_2$H$_4$ in the \textbf{Upper} and \textbf{Lower} panel respectively. The absolute difference of the HF ground state energy with respect to the VQE ground state energy ($\color{black} -\CIRCLE-$) and with respect to the exact energy ($\color{red}-\CIRCLE-$) are plotted in the \textbf{Left} part. The ground state energies ($\color{black} -\star-$) and the first ($\color{blue} -\CIRCLE-$), second ($\color{red} -\blacksquare-$), and third ($\color{GPgreen} -\blacktriangle-$) excited state energies obtained with VQE+Q-SCEOM and exact diagonalization of the Hamiltonian are shown in the \textbf{Middle} and \textbf{Right} part respectively. The SSP method was used in the Q-SCEOM calculations.}}
\label{fig:c2h4_sceom}
\end{figure*}

Consequently, we look at the excited states of C$_2$H$_4$ using the Q-SCEOM method. The UCCSD ansatz, optimized by VQE for the ground state, was used for $U(\vec{\theta}_{\mathrm{opt}})$ in Eq.~\ref{eq:sceom_energy}. We also used particle and spin conserving singles and doubles excitation operators. We compare the efficiency of the GR and SSP methods for the construction of the off-diagonal elements of the $M$ matrix (see Sec.~\ref{sec:qsceom} for details). In Fig.~\ref{fig:gates_mats} we present the total number of gates (upper panel) and the total number of 2-qubit gates (lower panel) (the circuits were compiled with the H-series emulator) for each element of the $M$ matrix for the C$_2$H$_4$ molecule using the 8-qubit active space with torsion angle of 90\textdegree. More specifically, the circuits correspond to $U(\vec{\theta}_{\mathrm{opt}})\hat{G}_J\left|\psi_{\mathrm{HF}}\right>$ for the diagonal elements and $U(\vec{\theta}_{\mathrm{opt}})\left(\hat{G}_I +\hat{G}_J\right)\left|\psi_{\mathrm{HF}}\right>$ for the off-diagonal elements. There are 26 excitation particle and spin conserving operators. Therefore the $M$ matrix consists of 676 elements. Only the diagonal and the upper-half part of the matrix were calculated since $M$ is a symmetric matrix. The GR method was used for the plots in the left side while the SSP method was used for plots in the right side. It is evident that the SSP method reduces significantly the number of gates. This is in accordance to results presented in Sec.~\ref{sec:results_gs} for the calculation of ground state energies. We note here that in the case of the GR method, the structure of the matrices shown in Fig~\ref{fig:gates_mats} is directly analogous to the Hamming distance of the associated states to be combined. In particular, the improvement of the SSP method over the GR method becomes greater for values of the Hamming distance larger than 4, for which the GR method requires the construction exemplified in Fig. \ref{fig:triple}. (see Sec.~\ref{app:sec2} in the Appendix for more details).

Additionally, the gate decompositions of the multiconfiguration states for Hamming distances $\leq$ 4 also contribute to the reduced circuit sizes obtained from the SSP for the $M$ matrix elements. This is exemplified in Fig.~\ref{fig:sceom_circs}, where the GR and SSP circuits are shown for the construction of the $\left|\psi_{I+J}\right>$ state for the $M_{20}$ element for the Q-SCEOM calculation of the 8-qubit C$_2$H$_4$ at 90\textdegree\ torsion. We observe that 3 2-qubit entangling gates are needed for the SSP method, whereas the GR method requires 14 (see also Fig. \ref{fig:gr2_gr4}). 

Next, we look at the ground state and excited energies of C$_2$H$_4$ at various torsion angles. We present in Fig.~\ref{fig:c2h4_sceom} the results for the 4-qubit and 8-qubit case in the upper and lower parts, respectively. In the left panel we plot in red the correlation energy of the ground state energy as the absolute difference of the HF energy with respect to the exact result (first eigenvalue obtained with diagonalization of the Hamiltonian). As expected, the degree of correlation increases as we approach the 90\textdegree \ torsion. We also plot in black the absolute difference of the HF energy with respect to the VQE result obtained through optimization of the UCCSD ansatz. It is evident that at 90\textdegree \ torsion angle, VQE yields a higher energy compared to the exact result. We attribute this to the fact that the VQE optimization results in an unstable singlet state whereas the stable solution is a triplet, as discussed in Sec.~\ref{sec:results_gs}. Note here, that for all the VQE calculations reported in this section, prior to the calculation of the $M$ matrix, we limited our search to open-shell singlet states. Consequently, we plot in the middle panel the VQE result for the ground state energy and the energy of the first three excited states obtained with Q-SCEOM. We employed the SSP method for the construction of the off-diagonal elements of the $M$ matrix for C$_2$H$_4$ at various torsion angles. We compare the Q-SCEOM results to the eigenvalues obtained through diagonalization of the Hamiltonian. For both active spaces, Q-SCEOM reproduces qualitatively the exact result. Interestingly, at the 90\textdegree\ angle, Q-SCEOM recovers the correct total energy (which matches the exact result) even though VQE (which affects the Q-SCEOM result through the optimized ansatz that is used for the calculation of the $M$ matrix) yields a higher energy.

\section{Discussion}\label{discussion}

In this work, quantum circuit preparation of multiconfigurational states for quantum chemistry has been demonstrated and compared as the state preparation step for a range of applications, including ground state energy calculations using VQE, QCM, and QPE, as well as excited state energies calculated using Q-SCEOM. The use-case to demonstrate these techniques is twisted C$_2$H$_4$, whose energy surface passes through a strongly correlated point as a function of torsion angle. 

Multiconfigurational state preparation allows initial states that are more accurate than the single configuration HF state, which for QCM can facilitate a reduction in measurement overhead at a relatively small cost of extra circuit depth for the initial state, whereas for QPE the probability of accurately measuring the phase can be increased by a multiconfigurational initial state (relative to HF) since this probability is proportional to the overlap between the initial state and true ground state \cite{nielsenchuang11}. For Q-SCEOM, the off-diagonal elements of the $M$ matrix can be generated using multiconfigurational state preparation, hence the latter facilitates a framework for automatically building and running the Q-SCEOM algorithm for a given molecule.

Overall, we observe significantly more efficient circuits using the SSP method compared to the GR method. This is largely due to the utilization of the sparsity of chemical states that can be exploited by the SSP, in addition to the gate decomposition of particle-conserving GRs and their required external controls leading to larger gate and depth overheads for the GR method. A conceptual advantage of the GR method can be seen when considering the state preparation as transformations of a reference state; setting the rotation angle of all GRs to 0 is guaranteed to result in the state $|x_1\rangle$ where $x_1$ is the first ON configuration in the ordered input set. Since $x_1$ can be chosen from HF, each GR can be considered as an excitation of the HF reference, preserving familiar notions from classical computational chemistry. In the context of variational searches of the ground state energy, this property of the GR method prevents unpredictable local minima at small values of variational parameters (which the SSP method can be susceptible to), and may be beneficial for analyzing contributions of specific basis states to the total electronic correlation.

In terms of future applications, our implementation easily allows for the composition of different ansatz circuit structures. For example, appending generalized \cite{lee19} UCC unitaries to a multiconfigurational state circuit prepared using the GR or SSP methods provides a framework for multireference methods \cite{lyakh12} at the quantum circuit level, such as multireference UCC. We also note a recent work that extended the reference state error mitigation scheme \cite{lolur23} to multireference states \cite{zou25} using GRs. An interesting future direction would be to apply the SSP method to prepare the multireference states used for error mitigation, with the potential of significant reduction in circuit resources. 

In addition, since multiconfigurational state preparation as presented here can be seen as an approach of loading classical (chemical) data to a quantum processor and preparing a quantum state to represent this data, this could be compared to preparation of multi-determinant states using quantum read-only memory (QROM) \cite{tubman18}, or to sparse state preparation using quantum random access memory \cite{deveras22}. We note that while these techniques achieve a similar asymptotic scaling to the SSP method in the number of 2-qubit gates ($O(Dn_q)$, our implementation of multiconfigurational state preparation does not require controlled operations between auxiliary qubits and state qubits. We also note a recent work \cite{berry25} that utilizes QROM to reduce the Toffoli count for matrix product state preparation.

In conclusion, this work provides a framework for preparing quantum circuit representations of multiconfigurational states, as implemented in \mbox{InQuanto} \cite{inquanto_web, inquanto_medium, inquanto_docs}. Overall, the impact of this work can be summarized by the following points. 
\begin{enumerate}
\item[i)] We provide a novel implementation for the use of externally controlled GRs for multiconfigurational state preparation, which automatically finds the required external controls of GRs for a given ordered set of ON configurations. 
\item[ii)] While the SSP algorithm was published in a previous article \cite{gleinig21}, this work demonstrates novel applications of the SSP method for a range of algorithms currently proposed for studying the ground and excited states of molecules quantum computationally, i.e. VQE, QCM4, QPE, and Q-SCEOM. 
\item[iii)] Specific to the application of the SSP method to VQE, our implementation allows for the use of SSP to generate a variational ansatz, in which the lowest energy state vector can be obtained by optimizing the angles of the 1-qubit unitaries within the SSP circuit.
\end{enumerate}
The utility of multiconfigurational initial states is shown for various quantum computational approaches to quantum chemistry. In particular, when the state to be prepared exhibits sparsity in the sense of $D << 2^{n_q}$, very efficient circuit representations can be achieved, therefore boosting the accuracy of QPE, for example, or reducing the measurement overhead of quantum subspace methods, without a large overhead in circuit depth. In addition, multiconfigurational state preparation is useful for enabling certain methods (such as Q-SCEOM) which require circuit constructions of selected excitations of a reference state, thus enhancing the tool set for quantum approaches to ground and excited states of molecules.

\vspace{10mm}

\section*{Author contributions}
All authors contributed to the conceptualization of the work. G.G.D, G.P., and D.Z.M. contributed to the implementation of methods and analysis of results. All authors contributed to the manuscript.

\section*{Conflicts of interest}
\noindent The authors declare no conflict of interest.

\section*{Data availability}
\noindent The data and binary version of the software supporting the conclusions of this paper are available from the corresponding author upon request. Any provision of the software or data is at the discretion of the providing company. A trial version of InQuanto is available at \url{https://www.quantinuum.com/products-solutions/inquanto-trial}. InQuanto version 5.0 (\url{https://inquanto.quantinuum.com/index.html}) is used in this work. Pseudocode for the GR method is presented in this article (see Algorithms \ref{alg:gr} and \ref{alg:lambda_gt_2}). See also pseudocode for the SSP method reported in the original paper by Gleinig and Hoefler (\url{https://doi.org/10.1109/DAC18074.2021.9586240}). A repository is available that contains representative starting datasets and notebooks to allow similar conclusions to be drawn to those of the manuscript based on the workflow in the notebooks and self-written replacements for calls to proprietary code (\url{https://doi.org/10.5281/zenodo.17466867}).

\section*{Acknowledgments}
\noindent We thank Alec Owens and Carlo Gaggioli for helpful feedback and comments. We also acknowledge Josh J. M. Kirsopp  and Iakov Polyak for useful discussions.



\balance


\bibliography{rsc} 
\bibliographystyle{rsc} 

\onecolumn
\newpage
\section*{Appendix}
\appendix
\counterwithin{figure}{section}
\section{Circuit decompositions for GR Method} \label{app:sec1}

\begin{figure*}[ht]
    \includegraphics[width=1.0\linewidth]{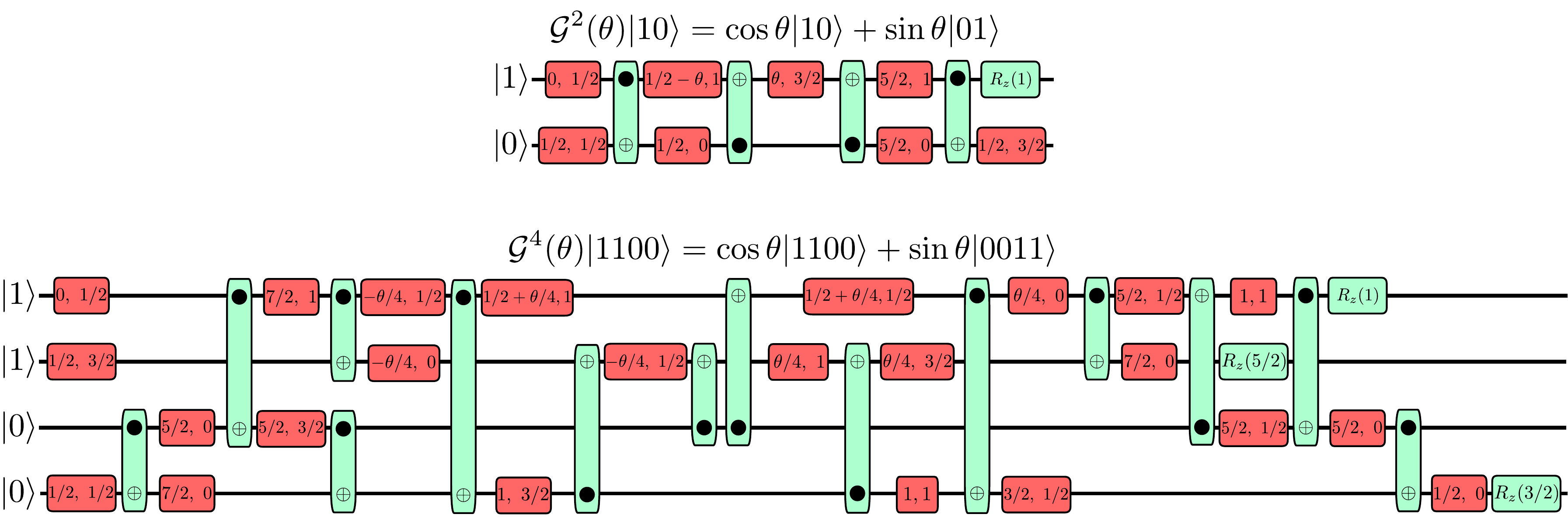}\hfill
\caption{Circuits for 1-body (2-qubit) $\mathcal{G}^2$ and 2-body (4-qubit) $\mathcal{G}^4$ rotations, where state vectors are obtained up to global phase. Green vertical bars represent the ZZMax 2-qubit entangling gate, while red horizontal bars represent PhasedX($\alpha, \beta) = R_z(\beta)R_x(\alpha)R_z(-\beta)$ 1-qubit rotations \cite{h11}. All gate angles are in units of $\pi$. Subscripts on $\mathcal{G}^2$ and  $\mathcal{G}^4$ are omitted.}
\label{fig:gr2_gr4} 
\end{figure*}

\begin{figure*}[ht]
\centering
\includegraphics[width=0.5\linewidth]
{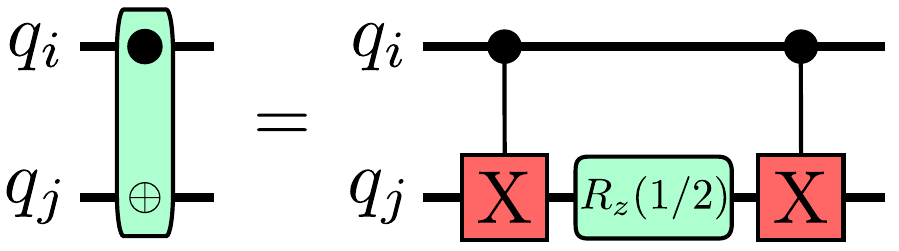}
\caption{The 2-qubit entangling gate ZZMax \cite{h11} in terms of CNOT and $R_z$ gates. The equality is up to global phase. Gate angles are in units of $\pi$.}
\label{fig:zzmax}
\end{figure*}

\begin{figure*}[ht]
\centering
\includegraphics[width=0.4\linewidth]{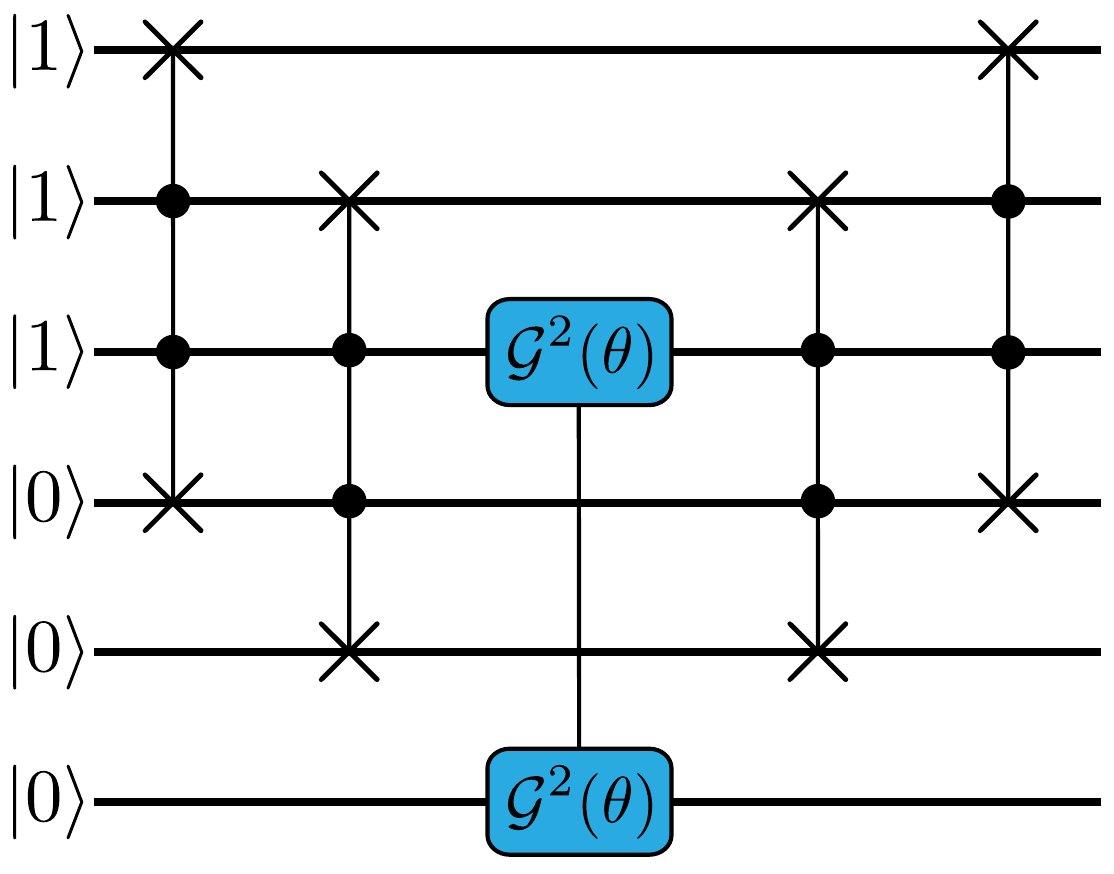}
\caption{Circuit to perform a ($\lambda=3$)-body excitation of the ON configuration $\ket{111000}$, yielding $\cos{\theta}\ket{111000} + \sin{\theta}\ket{000111}$, where $\theta$ is in units of $\pi$. The subscript $e$ in $\mathcal{G}^2_e$ is omitted in the figure. See top panel of Fig. \ref{fig:gr2_gr4} for the decomposition of the central $\mathcal{G}^2$ in the H-series gate set\cite{h11}.}
\label{fig:triple}
\end{figure*}

In Fig. \ref{fig:gr2_gr4} we show the gate decompositions of the 1-body ($\mathcal{G}^2$) and 2-body ($\mathcal{G}^4$) rotations. Compared to the corresponding circuits shown in previous work \cite{arrazola22}, we observe a similar number of 2-qubit entangling gates (ZZMax\cite{h11} here and CNOTs in Arrazola \textit{et al.} \cite{arrazola22}). The Hadamard gates of the decompositions shown in \cite{arrazola22} are absorbed into the PhasedX($\alpha, \beta) = R_z(\beta)R_x(\alpha)R_z(-\beta)$ \cite{h11} 1-qubit rotations, and additional $R_z$ and PhasedX rotations are required to achieve a 2-qubit operation equivalent to a CNOT, resulting in differences in the total number of 1-qubit gates. For completeness, the action of the 2-qubit ZZMax is depicted in Fig. \ref{fig:zzmax} in terms of CNOTs and $R_z$.

This section also describes our scheme to build the excitation gadget that connects basis states separated by $\mathfrak{h}(x_1, x_d) > 4$ Hamming distance ($\lambda>2-$body excitation), when preparing the multiconfigurational state using the GR method. Unlike Fig. 5 of in Arrazola \textit{et al.} \cite{arrazola22}, here each SWAP is not controlled on all other qubits in the circuit, but only on ``minority'' qubits, i.e. those qubits with minority binary values (see $q_{\text{minor}}$ in Algorithm \ref{alg:lambda_gt_2}). This results in lower overhead of external controls that scale with the number of minority qubits $n_{\text{minor}}$ rather than $n_q$ as in in Arrazola \textit{et al.} \cite{arrazola22}. However, this leads to the possibility that \textit{i}) the control states (conditions of the external control) of a SWAP are matched by the qubits of one of the previous basis states, or \textit{ii}) a swapped version of the first basis state ($x_1'$) becomes equal to one of the previous basis states (note that qubits of $x_1'$ at a given iteration of the WHILE loop lines 5 - 13 of Algorithm \ref{alg:lambda_gt_2} form the control states for the next iteration). If \textit{i}) or \textit{ii}) occurs, it is not necessarily a problem once the central $\mathcal{G}^1$ is also controlled on the minority qubits of (``fully swapped'') $x_1'$. The latter implies that the external controls of the central $\mathcal{G}^1$ also scale with $n_{\text{minor}}$, while a cheaper alternative would be to use the usual $\mathcal{G}^1$ controls from Algorithm \ref{alg:gr}. However, in order for the external controls found by Algorithm \ref{alg:gr} to be usable for the central $\mathcal{G}^1$ of the $\lambda>2$ excitation gadget, we find that both \textit{i} and \textit{ii} must not occur.

\begin{algorithm}[hbt!]
\caption{GR for $\mathfrak{h}(x_1, x_e) > 4$}\label{alg:lambda_gt_2}
\textbf{Input:} Ordered set of bit strings $(x_1, ..., x_e)$. \\

\textbf{Output:} Sequence of \texttt{SWAP} gates, central $\mathcal{G}^2_e$, and their external controls. \\

\nl $q_{\text{minor}} \gets
    \begin{cases}
        1 & \text{if $n_{\text{elec}} \leq \lfloor n_q/2 \rfloor$} \\
        0 & \text{otherwise}
    \end{cases}
    $ \ \ \;

\nl Make a copy of $x_1$, labeled $x_1'$\;

\nl Initialize ordered \texttt{dict} \texttt{ctrlSWAPs}$=\{\}$, and \mbox{$\texttt{k\_matches\_ref}$ $\gets$ FALSE}

\nl \While{$\mathfrak{h}(x_1', x_e) > 2$}{
\nl     Find lowest $i$ such that $x_1'(i) = 1 \neq x_e(i)$, and lowest $j$ such that $x_1'(j) = 0 \neq x_e(j)$. Store \texttt{SWAP} indexes $(i, j)$\;
\nl Initialize empty ordered set $\textbf{\textit{i$'$}}=\{\}$\;
\nl     \For{$i' \notin \{i, j\} = 1$ \KwTo $n_q$}{
\nl         \If{$x_1'(i') = q_{\text{\normalfont minor}}$}{
\nl             Append external control index $i'$ to $\textbf{\textit{i$'$}}$, with control state $q_{\text{minor}}$.
        }
    }
\nl Append (key, value) pair ($(i, j)$, $\textbf{\textit{i$'$}}$) to \texttt{ctrlSWAPs}. Swap $i, j$ of $x_1'$, reduce $\mathfrak{h}(x_1', x_e)$ by 2\;
\nl     \For{$k=2$ \KwTo $e-1$}{
\nl         \If{$x_1'$ = $x_k$}{
\nl             Set $\texttt{k\_matches\_ref}$ $\gets$ TRUE\;
        }
    }
}
\nl Use the remaining 2 nonzero indexes of $((x_1'(1) \oplus x_e(1), ..., x_1'(n_q) \oplus x_e(n_q))$ to define the central $\mathcal{G}^2_e$. Initialize \mbox{$\texttt{allow\_G\_ctrls}$ $\gets$ TRUE}\;
\nl Use Algorithm \ref{alg:gr} lines 4-6 to find (potential) external control indexes $\textbf{\textit{i$_p$}}$ of $\mathcal{G}^2_e$\;
\nl \If{\normalfont $\texttt{k\_matches\_ref}$}{
\nl     \For{\normalfont $\textbf{\textit{i$'$}}$ in \texttt{ctrlSWAPs}}{
\nl         \If{$\textbf{\textit{i$_p$}} \subseteq \textbf{\textit{i$'$}}$ \bf{and} \normalfont $x_1'(i_p) = q_{\text{\normalfont minor}} \forall i_p \in \textbf{\textit{i$_p$}}$} {
\nl             $\texttt{allow\_G\_ctrls}$ $\gets$ FALSE\;
        }  
    }
}
\nl     \If{\normalfont $\texttt{allow\_G\_ctrls}$}{
\nl         Set external control indexes of $\mathcal{G}^2_e$ to $\textbf{\textit{i$_p$}}$, with control states $x_1'(\textbf{\textit{i$_p$}})$\;
    }
\nl     \Else{
\nl         Use lines 7 - 9, in which the $(i, j)$ are now the indexes of the central $\mathcal{G}^2_e$, to find external controls of $\mathcal{G}^2_e$\;
}
\nl Return \texttt{ctrlSWAPs}, indexes of $\mathcal{G}^2_e$, and $\mathcal{G}^2_e$ external controls. 
\end{algorithm}

\section{Circuit Resources and Hamming Distances for the \texorpdfstring{$M$}{M} matrix in Q-SCEOM}\label{app:sec2}

\begin{figure*}[ht]
    \begin{minipage}[b]{\textwidth}
        \includegraphics[width=0.33\linewidth]{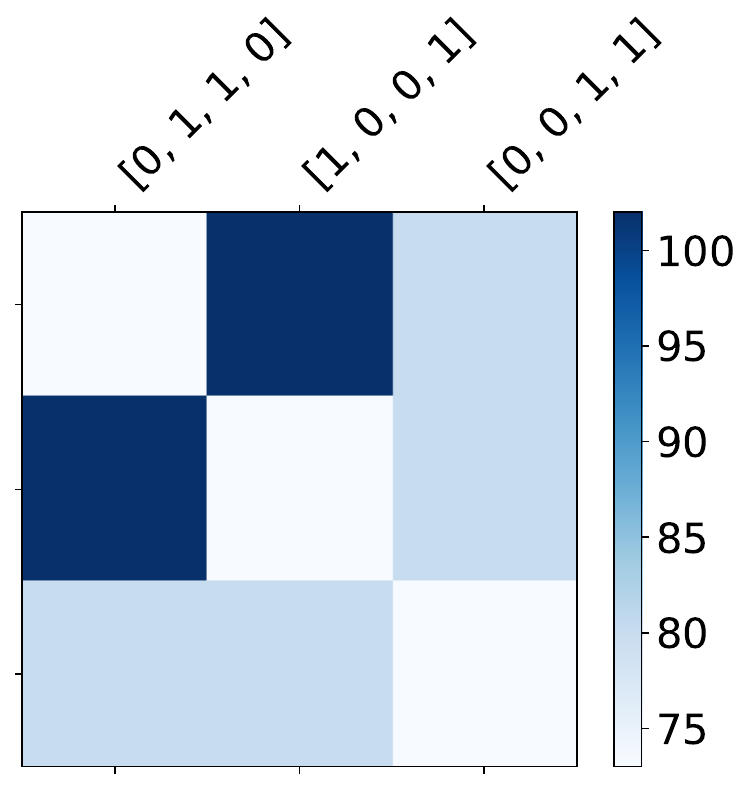}\hfill
        \includegraphics[width=0.33\linewidth]
        {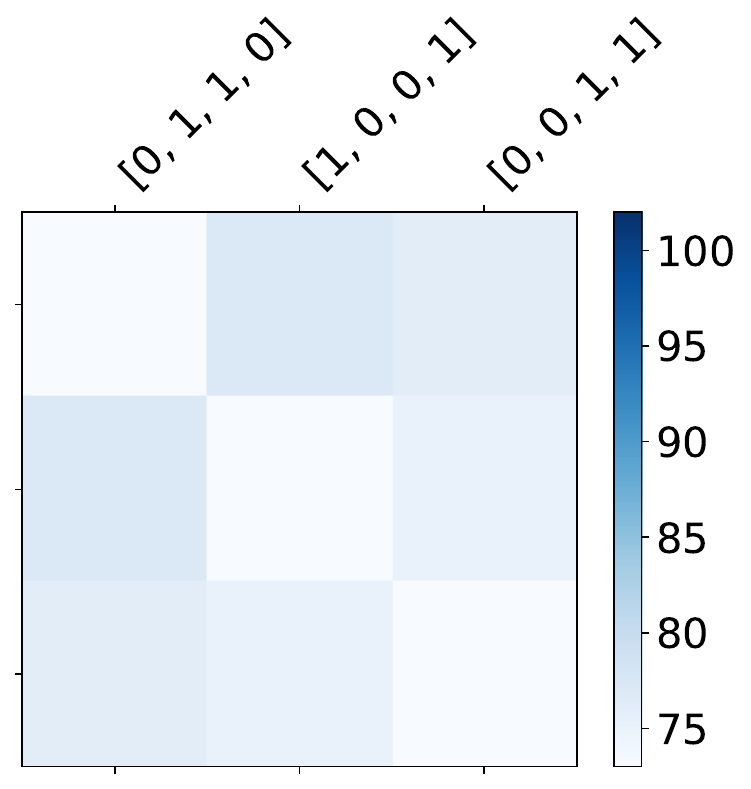}\hfill
        \includegraphics[width=0.30\linewidth]{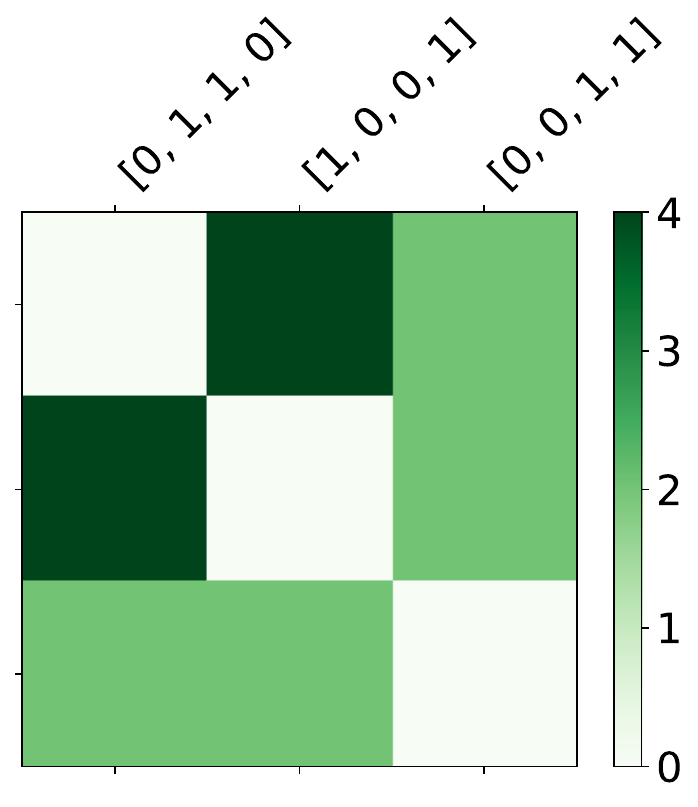}\hfill
        \caption{\small{Total number of gates using the \textbf{Left)} GR method, and the \textbf{Middle)} SSP method for the construction of the 'ket' states of the $M$ matrix for the 4-qubit active space for C$_2$H$_4$ at 90 \textdegree torsion. \textbf{Right)} The hamming distance between the $I$, $J$ states that are combined  using the previous methods to construct the $\left|\psi_I +\psi_J\right>$ state. The $\hat{G}_I\left|\psi_{\mathrm{HF}}\right>$ states for each element are shown as x tick labels.}}
        \label{fig:appx_mats}
    \end{minipage}
\end{figure*}

\begin{figure*}[ht]
\centering
        \includegraphics[width=0.33\linewidth]
        {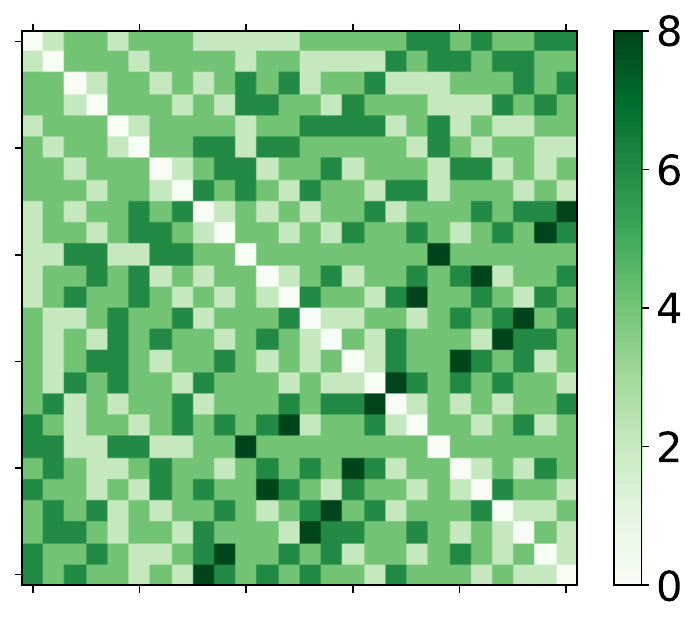}\hfill
        \caption{\small{Hamming distance between the $I$, $J$ states that are combined to construct the $\left|\psi_I +\psi_J\right>$ states for C$_2$H$_4$ using the 8-qubit active space.}}
        \label{fig:appx_mats2}
\end{figure*}

In this section, we plot the Hamming distances between the $\left|\psi_{I+J}\right>$ and $\left|\psi_{\mathrm{HF}}\right>$ states, used to build the $M$ matrix for Q-SCEOM. The Hamming distances have a significant impact on the circuit resources used for each element of the $M$ matrix, particularly for the GR method. Fig. \ref{fig:appx_mats} corresponds to the 4-qubit active space, where the left and middle panels show the total number of gates for the GR and SSP methods, respectively, while the right panel shows the Hamming distances. Fig. \ref{fig:appx_mats2} shows the Hamming distances for the 8-qubit active space.

\section{Plots of Orbitals}

\begin{figure*}[ht]
    \begin{minipage}[b]{\textwidth}
        \includegraphics[width=0.9\linewidth]
        {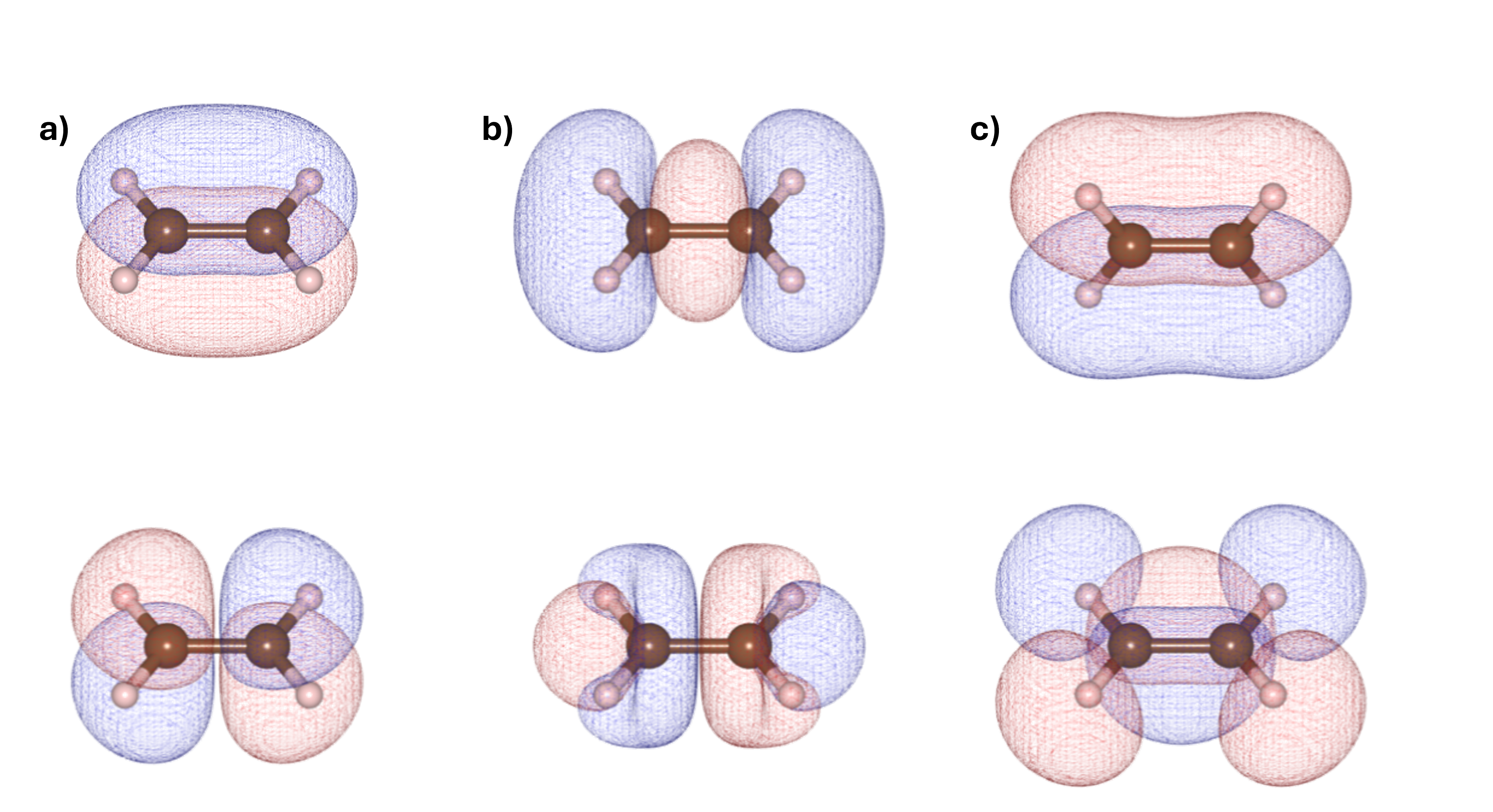}\hfill
        \caption{\small{UHF spatial orbitals of C$_2$H$_4$ at 0\textdegree torsion angle that were used for the construction of the active spaces. The HOMO and LUMO shown in \textbf{a)} were used for the 4-qubit case. The 8-qubit case was augmented with the orbitals shown in \textbf{b)} and the 12-qubit case used all six orbitals shown in the figure. Occupied and virtual orbitals are shown in the \textbf{upper} and \textbf{lower} part respectively. The isosurface level was set to 0.02. \texttt{PySCF}\cite{pyscf} was used for the HF calculation.}}
    \label{app:orbs_plot}
    \end{minipage}
\end{figure*}

In this section we plot the spatial orbitals used in the active spaces of C$_2$H$_4$, shown in Fig. \ref{app:orbs_plot}.

\end{document}